 \documentclass[lettersize,journal]{IEEEtran}
\usepackage{amsmath,amsfonts}
\usepackage{hyperref}
\usepackage{setspace}
\usepackage{amssymb}
\usepackage{algorithmic}
\usepackage{bm}
\usepackage{algorithm}
\usepackage{array}
\usepackage[caption=false,font=normalsize,labelfont=sf,textfont=sf]{subfig}
\usepackage{textcomp}
\usepackage{stfloats}
\usepackage{xcolor}
\usepackage{color}
\usepackage{url}
\usepackage{verbatim}
\usepackage{graphicx}
\usepackage{epstopdf}
\usepackage{cite}
\usepackage{multirow}
\usepackage{changepage}
\newtheorem{prop}{Proposition}[section]
\newtheorem{defi}{Definition}[section]

\newtheorem{remark}{Remark}

\begin{document}

\title{Neural Quantile Optimization for Edge-Cloud \textcolor{black}{Networking}}

\author{
Bin Du$^1$\thanks{$^1$School of Mathematical Sciences, Peking University, Beijing 100871, China (dubinpku@pku.edu.cn)},
\and
He Zhang$^2$\thanks{$^2$Beijing International Center for Mathematical Research, Peking University, Beijing 100871, China (zhanghe@bicmr.pku.edu.cn)},
\and
Xiangle Cheng$^3$\thanks{$^3$Huawei, Beijing 100095, China (chengxiangle1@huawei.com)},
\and
Lei Zhang$^{4*}$\thanks{$^{4}$Beijing International Center for Mathematical Research, Center for Machine Learning Research, Center for Quantitative Biology, Peking University, Beijing 100871, China (zhangl@math.pku.edu.cn)}\thanks{$^*$Corresponding author, L.Z. was supported by the National Natural Science Foundation of China (No.12225102, T2321001, and 12288101) }
}

% <-this % stops a space
% <-this % stops a space

%\markboth{Journal of \LaTeX \ Class Files,~Vol.~14, No.~8, August~2021}%


\maketitle

\begin{abstract}
We seek the best traffic allocation scheme for the edge-cloud \textcolor{black}{networking subject to SD-WAN architecture} and burstable billing. First, we formulate a family of quantile-based integer programming problems for a fixed network topology with random parameters describing the traffic demands. Then, to overcome the difficulty caused by the discrete feature, we generalize the Gumbel-softmax reparameterization method to induce an unconstrained continuous optimization problem as a regularized continuation of the discrete problem. Finally, we introduce the Gumbel-softmax sampling neural network to solve optimization problems via unsupervised learning. The neural network structure reflects the edge-cloud \textcolor{black}{networking} topology and is trained to minimize the expectation of the cost function for unconstrained continuous optimization problems. The trained network works as an efficient traffic allocation scheme sampler, outperforming the random strategy in feasibility and cost value. Besides testing the quality of the output allocation scheme, we examine the generalization property of the network by increasing the time steps and the number of users. We also feed the solution to existing integer optimization solvers as initial conditions and verify the warm-starts can accelerate the short-time iteration process. The framework is general, and the decoupled feature of the random neural networks is adequate for practical implementations.

 \end{abstract}


\section{Introduction}
\label{sec:intro}

{\color{black}
\IEEEPARstart{w}{ith} the advent of the digital age, there is a need for high-speed transmission of massive data over Wide Area Networks (WANs). This poses higher requirements for the transmission quality and capacity of WANs. For instance, efficient and secure data communication methods are needed between the headquarters of a company and its various branch offices. Subsidiaries consolidate their computing resources to upload data to centralized cloud computing service centers and obtain services from the cloud \cite{yang2019software}. Software-Defined Wide Area Networking (SD-WAN) is an automated programming approach used to manage enterprise network connectivity and circuit costs. It extends Software-Defined Networking (SDN) into applications that enable the rapid creation of intelligent hybrid WANs \cite{SD-WANdefine}. In traditional networks, packet processing is primarily performed based on a single or a few attributes of packets, such as the longest destination IP prefix, destination media access control (MAC) address or IP address, and port numbers of Transmission Control Protocol (TCP) and User Datagram Protocol (UDP). SDN allows us to manage traffic based on more attributes of packet headers through the Control-Data-Plane Interface (CDPI), such as the OpenFlow protocol \cite{Traditionalwan}.  

\begin{figure}
    \centering
    \includegraphics[width=0.99\linewidth]{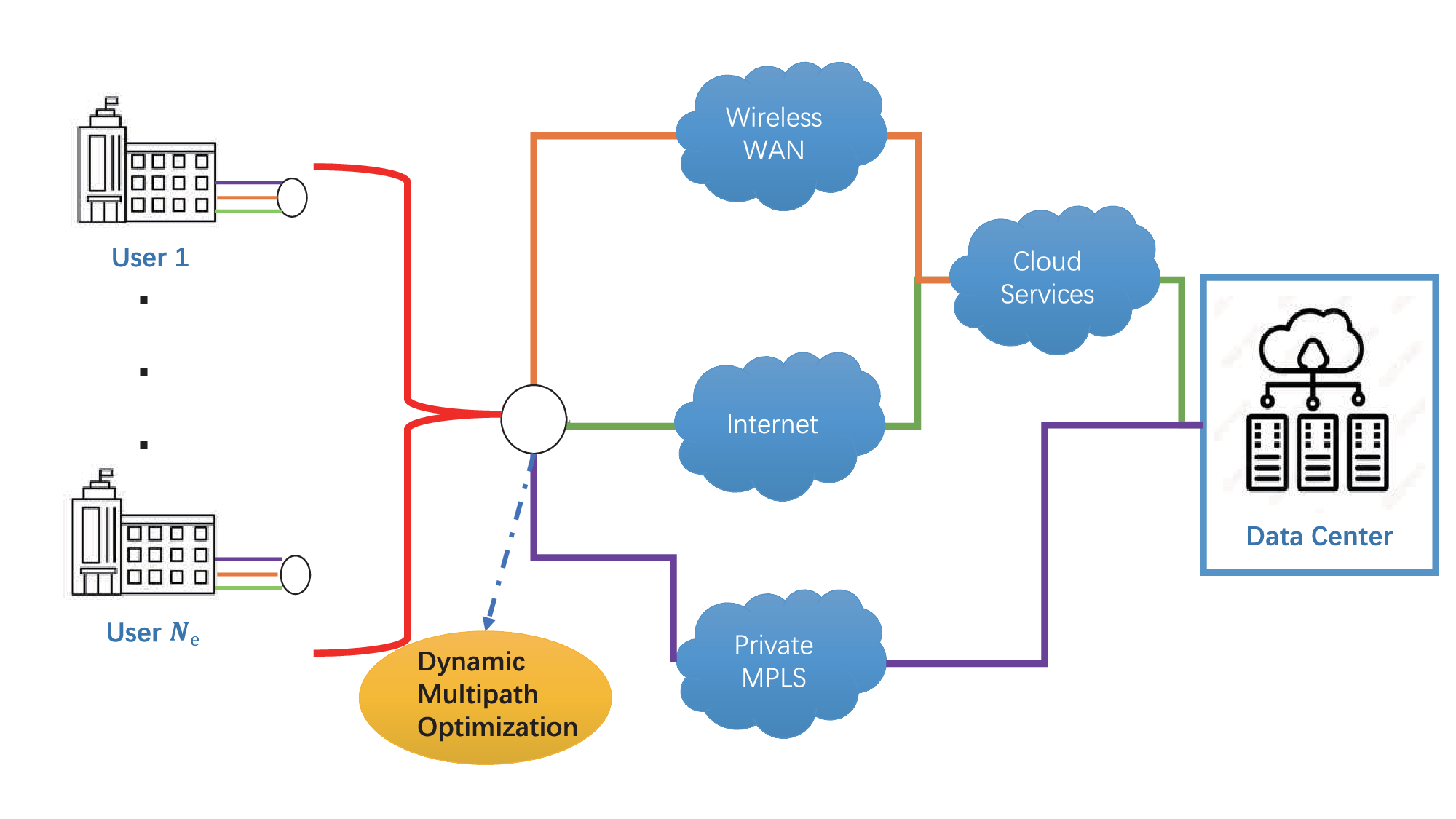}
    \caption{The architecture of SD-WAN consists of business-grade IP VPN, broadband Internet, and wireless services.\cite{SD-WANdefine}}
    \label{fig:SD-WAN}
\end{figure}

SD-WAN concerns itself with the endeavor of intelligently crafting network scheduling strategies, thereby bestowing upon users more convenient and cost-effective network services\cite{SD-WANdefine,Singh21}. Figure~\ref{fig:SD-WAN} displays the architecture of SD-WAN. To efficiently manage applications, highlighted in orange, traffic is dynamically optimized across the most appropriate WAN path through multipath optimization. Our work discusses the network scheduling problem in SD-WAN.
}

The network scheduling problem has been a research hot spot in network flow optimization in recent years. The greedy strategies are common approaches to finding allocation schemes for real-time task scheduling\cite{meng2019dedas}. Another way to solve network scheduling is to mark the bandwidth allocation of tasks by binary variables. The network scheduling of edge computing can be modeled as a class of constrained integer programming problems with input parameters\cite{mao2017survey,2020Safeguard,optimalnetwork}. Characterized by their discrete search spaces, solving the constrained integer programming is normally NP-hard \cite{papadimitriou1981complexity, 2010Percentile,jalaparti2016dynamic,kotary2021learning}. Many algorithms have been developed for integer programming problems, such as traditional greedy algorithm\cite{shenmaier2003greedy}, evolutionary algorithm \cite{ben1997genetic}, exact algorithms represented by branch and bound and cutting plane methods\cite{clausen1999branch, MIPbook}. Among them, based on the theory of precise algorithms, researchers have developed software, such as SCIP\cite{BestuzhevaEtal2021OO}, CPLEX\cite{cplex2009v12}, and Gurobi\cite{Gurobi}. For specific problems, the performance of the solver depends heavily on the initial guesses \cite{berthold2006primal}. 

{\color{black}

The online feature of network scheduling requires solving integer programming problems within a limited time frame, which depends on the billable bandwidth. Consequently, given a time limit, the desirable approach should be capable of identifying the best possible solution (may not be the optimal solution) before reaching the limit. However, when dealing with the problem at hand, commercial solvers that employ accurate methods often need a considerable amount of time to generate the optimal solution. Naively truncating the solving process by time constraints frequently leads to an infeasible solution.} To address this issue and expedite the generation of high-quality initial solutions for discrete edge-cloud traffic scheduling problems, we develop a neural network that employs sampling techniques to produce feasible solutions of superior quality rapidly.

In this work, the pricing scheme of our edge-cloud traffic scheduling problem considers the $95^{\text{th}}$ percentile billing method commonly used in industry standards,\cite{95thpercentile} and the different traffic requirements of several edge-devices compete for limited bandwidth resource allocation. We model the allocation selection as a class of constrained integer programming problems with the traffic demand and the link capacities as input parameters. Since the objective function is a piecewise linear function of the billable bandwidth depending on the $95^{\text{th}}$ percentile of the bandwidth distribution over a monthly time period, the resulting problem belongs to the field of quantile optimization, which adds extra computational complexity to the optimization problems but benefits the stability \cite{2021Quantile}.

\textcolor{black}{We improve the WAN Egress Traffic Allocation Model proposed in \cite{Singh21,heng_mccoll_2021} by considering more practical network topology and constraints. In order to align with the SD-WAN architecture depicted in Figure~\ref{fig:SD-WAN},} the network topology in our model includes two layers, and the objective function also includes the cost generated by the link between the hub and internet service providers, which couples the edges with each other. Thus, the resulting constrained optimization problem also describes the competition among the traffic demands from different edges. 

\textcolor{black}{The enhancement of solving efficiency for integer programming problems using neural networks has always been a pivotal research topic. In the existing literature, e.g., \cite{NEURIPS2021,nair2020solving,zhang2023survey}, researchers often preselect specific mixed integer programming (MIP) problems of interest and then design neural networks using supervised or reinforcement learning approaches based on the problem characteristics. Traditional exact solvers for integer programming employ techniques such as branch and bound and cutting plane methods to systematically generate a search tree by selecting suitable variables, aiming to reduce the primal-dual gap. Machine learning-based methods can be broadly categorized into two directions: techniques based on exact solvers and approaches approximating solutions using heuristics. The former revolves around providing better variable selection orders for algorithms like branch and bound in exact solvers to assist in solving MIPs. However, due to the NP-hard nature of MIPs, techniques relying on solvers often consume time and resources when generating problem solutions. The latter approach mostly employs supervised learning to utilize exact solvers for solving MIPs and generating training data, enabling the neural network to learn how to generate high-quality feasible solutions. Similarly, as the MIP problem sizes increase, obtaining training data using exact solvers becomes exceedingly challenging. Indeed, when modeling MIP problems in edge-cloud networks, we often encounter scenarios where the number of users accessing the network varies. It is undesirable to train different neural networks for MIP problems with different user counts. Additionally, obtaining training data for large-scale problems can be expensive. {\color{black} Our goal is to design an unsupervised neural network model that does not depend on exact solvers. It should be trainable on small-scale data, capable of producing good initial guesses for the problem, and show a certain level of generalization for larger-scale problems.}
}

\subsection{Contributions and Organization of the Paper}

{\color{black} 

The paper presents a new machine-learning technique that can enhance or replace the traditional discrete method for solving edge-cloud networking issues to minimize cost expectations. As numerical simulations, we examine test problems that offer insights into real-world networking challenges. Compared to the baselines, our approach shows better scalability in terms of network topology and significantly faster performance at a large scale.} We highlight the following main contributions.

\begin{enumerate}
    \item We develop models for the network scheduling problem based on the $95^{\text{th}}$ percentile billing strategy. The resulting constrained integer programming problems correspond to the multipath optimization problem in SD-WAN with more general constraints, including fractional linear constraints.

    \item {\color{black} We apply the Gumbel-Softmax reparameterization technique \cite{2016Categorical,9729603, Concrete2017} to the discrete integer programming problem. Instead of solving the relaxation problem, we utilize continuous reparameterization to model the discrete decision variables using continuous neural networks.}
    
    \item We employ a neural network-based sampling strategy to generate feasible solutions. Additionally, we utilize unsupervised learning for training purposes. Unlike some existing approaches, the training and usage of our model are independent of the MIP exact solvers.

    \item We adopt a decoupling approach, allowing for parallel processing of different schemes across different time steps, users, and traffic types in the training and implementation of the sampling network.
 {\color{black}
    \item We test the proposed sampling network's performance in various aspects, including the generalization property. Based on the numeric results, our sampling network significantly outperforms the random baseline. It can serve as a preconditioning procedure for commercial solvers by using the samples as ``warmed-up" starts.
}
\end{enumerate}

The paper is organized as follows. Section~\ref{sec:softmax} covers the Mathematical preliminaries of our work. We review the Gumbel-Softmax reparameterization trick and introduce a general framework for approximating an integer programming problem by a series of continuous problems so that the later implementation of the random neural network is possible. The target constraint integer programming problem that models the edge-cloud network scheduling is introduced in Section~\ref{sec:problem}. We discuss how we use random neural networks to generate warm starts for the target problems in Section~\ref{sec:network}, while Section~\ref{sec:num} includes the corresponding numerical results. We close the paper with a brief summary and discussion in Section~\ref{sec:conclusion}.

\section{The Gumbel-Softmax Reparameterization}
\label{sec:softmax}

This section introduces a general framework for transforming integer programming problems into continuous optimization problems. In particular, given an integer optimization problem, a series of unconstrained continuous optimization problems are introduced whose minimizers approximate the underlying categorical solution asymptotically. We first absorb the constraints by the method of Lagrange multipliers\cite{donti2021dc3, wolsey2020integer}. Afterward, considering the decision variable as a random variable, we rewrite the cost as a function of the categorical distribution. Finally, we apply the Gumbel-Softmax reparameterization trick\cite{2016Categorical} as a sampling process whose parameter gradients can be easily computed to implement the backpropagation algorithm.

Consider a family of integer programming problems
\begin{equation}\label{eq:int_p0}
\min_{y\in \Omega} f(y;\theta), \quad \mathrm{s.t.}\; g(y;\theta) \leq 0, \; h(y;\theta)  = 0,
\end{equation}
where $\theta\in \Theta$ denotes the parameters of the program and $y$ corresponds to the decision variable. For any fixed parameter $\theta$, the constrained programming problem \eqref{eq:int_p0} poses a task of minimizing the objective function $f$, as a function of the decision variable $y$, subject to the condition that the constraints are satisfied. The domain of the decision variable, denoted by $\Omega$, is a finite discrete set encoded as a set of $d$-dimensional one-hot vectors lying on the corners of the $(d-1)$-dimensional simplex $\Delta^{d-1}$\cite{2016Categorical,Concrete2017}. In particular, denote $\Omega = \{y_1,y_2, \dots, y_{d}\}$, where $y_{i}$ is an one-hot vector such that $i^{\mathrm{th}}$-component is $1$ and others are zero. The equality and inequality constraints are represented by vector-valued functions $h$ and $g$ (potentially nonlinear and non-convex) in \eqref{eq:int_p0}, respectively. We denote $\Omega_{\theta}$ as the feasible set of the decision variable, that is, 
\begin{equation*}
\Omega_{\theta} = \{y \in \Omega \;|\; g(y;\theta)\leq 0,\; h(y;\theta) = 0\}.
\end{equation*}
For simplicity, we assume that $\Omega_{\theta}$ is nonempty for all $\theta\in \Theta$, and for any fixed problem parameter $\theta\in \Theta$, the objective function $f$ and each component of $g$ and $h$ in \eqref{eq:int_p0} are smooth functions of $y$ in $\Delta^{d-1}$. Since $d$ is the cardinality of the decision variable domain, it suffers from the curse of dimension in general. For example, if we consider the $0-1$ knapsack problem\cite{salkin1975knapsack} of $n$ objects, then the decision variable $y$ is encoded as a $2^{n}$-dimensional one-hot vector, while the parameter $\theta$ stores the information of the object's weights and values.

To begin with, by viewing the constraints in \eqref{eq:int_p0} as a form of regularization\cite{donti2021dc3}, we formulate the soft-loss function
\begin{equation}\label{eq:soft_loss}
f_{\mathrm{soft}} = f(y;\theta) + \lambda_{g} \|\mathrm{ReLU}(g(y;\theta)) \|^{2}_{2} + \lambda_{h} \| h(y; \theta)\|_{2}^{2},
\end{equation}
where $\lambda_{g}, \lambda_{h}>0$. The composite loss in \eqref{eq:soft_loss} contains objective and two penalty terms representing equality and inequality constraint violations. In general, we cannot apply gradient-based methods to find the optimizer of the soft-loss function $f_{\mathrm{soft}}(y;\theta)$ due to the discrete feature of $\Omega$. In \cite{chen2023monte}, the Monte Carlo Policy Gradient Method resolves the issue. Here, we approximate the integer optimization problem with a series of continuous problems to overcome the challenge.

We introduce a random variable $Y$ on $\Omega$ with a location parameter\cite{Concrete2017} $\bm{\alpha} = (\alpha_{1}, \alpha_{2}, \dots, \alpha_{d}) \in (0, +\infty)^{d}$ satisfying
\begin{equation}\label{eq:rvY}
\mathbb{P}(Y = y_{i}) = \frac{\alpha_{i}}{\|\alpha\|_{1}}, \quad \|\alpha \|_{1} = \sum_{i=1}^{d} |\alpha_{i}|,
\end{equation}
that is, after a normalization, $\bm{\alpha}/\|\bm{\alpha}\|_{1}$ corresponds to the probability mass vector of the random variable $Y$. Here, we do not impose $\bm{\alpha}$ satisfying $\|\bm{\alpha}\|_1 = 1$ since in Section~\ref{sec:network}, we connect the location parameter to the output of a random neural network, which is positive but not necessarily normalized. We have the following naive equivalence
\begin{equation}\label{eq:naive}
\min_{y\in \Omega} f_{\mathrm{soft}}(y; \theta) \Leftrightarrow \min_{\bm{\alpha}\in (0,+\infty)^{d}} \mathbb{E}\left[ f_{\mathrm{soft}}(Y; \theta)\right],
\end{equation}
where $\mathbb{E}[\cdot]$ denotes the expectation with respect to $(\Omega, \mathbb{P})$, that is,
\begin{equation}\label{eq:expect}
\mathbb{E}\left[ f_{\mathrm{soft}}(Y; \theta)\right] = \|\bm{\alpha}\|_{1}^{-1}\sum_{i=1}^{d} \alpha_{i} f_{\mathrm{soft}}(y_i; \theta).
\end{equation}
Thus, even though $\mathbb{E}\left[ f_{\mathrm{soft}}(Y; \theta)\right]$ is an explicit function of the continuous variable $\bm{\alpha}$, its evaluation requires knowing the value of $f(y;\theta)$ over the entire $\Omega$, that is, the continuous optimization problem in \eqref{eq:naive} shares the same computational complexity with the equivalent integer optimization problem. To avoid visiting the value of $f_{\mathrm{soft}}(y;\theta)$ over entire $\Omega$, we consider an empirical estimator of the expectation in \eqref{eq:expect}
\begin{equation}\label{eq:approx1}
\mathbb{E}\left[ f_{\mathrm{soft}}(Y; \theta)\right] \approx \frac{1}{N} \sum_{k=1}^{N} f_{\mathrm{soft}}(Y^{(k)}; \theta),
\end{equation}
where $\{Y^{(k)}\}_{k=1}^{N}$ are i.i.d. samples of the random variable $Y$ with location parameter $\bm{\alpha}$. We can also interpret the right-hand side of \eqref{eq:approx1} as the empirical loss function or empirical risk function\cite{kotary2021learning}. 

The Gumbel-Softmax reparameterization trick\cite{Concrete2017, 2016Categorical} provides a way to sample $Y^{(k)}$ based on the location parameter $\bm{\alpha}$. For $\tau >0$, we define the concrete random variable $X = (X_{1}, X_{2}, \dots, X_{d})\in \Delta^{d-1}$, given by
\begin{equation}\label{eq:repara}
X_{k} = \frac{\exp \left( (\log(\alpha_{k}) + G_{k})/\tau\right)}{\sum_{i=1}^{d}\exp\left( (\log(\alpha_{i}) + G_{i})/\tau\right)},
\end{equation}
where $G_{i}\sim \textrm{Gumbel}(0,1)$. The $\textrm{Gumbel}(0, 1)$ distribution can be sampled using inverse transform sampling by drawing $u_{i}$ from the uniform distribution on $[0,1]$ and taking $g_i = -\log(-\log(u_i))$. We review two useful properties of the concrete random variables\cite{Concrete2017}.

\begin{prop}\label{prop:gumbel} Let $X \sim \mathrm{Concrete}(\bm{\alpha}, \tau)$ with location parameter $\bm{\alpha} \in (0, +\infty)^{d}$ and temperature $\tau \in (0, +\infty)$, then for $k= 1,2,\dots, d$, we have
\begin{enumerate}
\item (Rounding) $\mathbb{P}(X_{k} > X_{i}, \; \forall i\not=k) = \alpha_{k}/\|\bm{\alpha}\|_1$,
\item (Zero temperature) $\mathbb{P} \left(\lim\limits_{\tau \rightarrow 0^{+}} X_{k}=1\right) =  \alpha_{k}/\|\bm{\alpha}\|_1.$
\end{enumerate}
\end{prop}

Compared with \eqref{eq:rvY}, Propositioin~\ref{prop:gumbel} confirms that, after rounding, the sample of the concrete random variable $X \sim \mathrm{Concrete}(\bm{\alpha}, \tau)$ follows the same distribution as $Y$, and the asymptotic behavior of  
the concrete random variable as temperature $\tau$ goes to $0^{+}$ is the same as rounding. Thus, let $\{X^{(k)}\}_{k=1}^{N}$ be i.i.d. samples of the concrete random variable $X \sim \mathrm{Concrete}(\bm{\alpha}, \tau)$, and we can rewrite the approximation in \eqref{eq:approx1} as
\begin{equation}\label{eq:approx2}
\mathbb{E}\left[ f_{\mathrm{soft}}(Y; \theta)\right] \approx \frac{1}{N} \sum_{k=1}^{N} f_{\mathrm{soft}}([X^{(k)}]; \theta),
\end{equation}
where $[X^{(k)}]$ denotes the rounding of $X^{(k)}$, that is,
\begin{equation*}
[X^{(k)}] = y_{j}\in \Omega \Leftrightarrow X^{(k)}_{j} > X^{(k)}_{i}, \; \forall i\not = j.
\end{equation*}
Notice that the domain $\Omega$ corresponds to the corners of $\Delta^{d-1}$, while $X$ is a random variable defined on $\Delta^{d-1}$. Therefore, from a geometric perspective, the rounding procedure seeks the corner of $\Delta^{d-1}$ nearest to the sample $X^{(k)}$.

The reparameterization in \eqref{eq:approx2} connects the objective function with the samples $X^{(k)}$, which smoothly depend on the location parameter $\bm{\alpha}$ according to \eqref{eq:repara}. We further drop the non-differentiable rounding procedure in \eqref{eq:approx2} and formulate the following unconstrained continuous optimization problem as the main result of this section.

\begin{defi}(Gumbel-Softmax reparameterization) Consider the integer programming problem in \eqref{eq:int_p0} with soft-loss function \eqref{eq:soft_loss}, given temperature $\tau>0$ and i.i.d. samples of $\mathrm{Concrete}(\bm{\alpha}, \tau)$ denoted as $\{X^{(k)}\}_{k=1}^{N}$, the corresponding Gumbel-Softmax reparameterization problem is
\begin{equation}\label{eq:p1}
\min_{\bm{\alpha} \in (0, +\infty)^{d}} \hat{f}_{\mathrm{soft}}^{N,\tau}(\bm{\alpha};\theta), \quad \hat{f}_{\mathrm{soft}}^{N,\tau} = \frac{1}{N} \sum_{k=1}^{N} f_{\mathrm{soft}}(X^{(k)}; \theta).
\end{equation}
Moreover, if $\hat{\bm{\alpha}}^{N, \tau}$ is a minimizer of the problem \eqref{eq:p1}, then we call the rounding of the normalized minimizer
\begin{equation}\label{eq:approx_min}
 \hat{y}^{N,\tau} =\left[\|\hat{\bm{\alpha}}^{N, \tau}\|^{-1}_{1} \hat{\bm{\alpha}}^{N, \tau}\right]   
\end{equation}
the Gumbel-Softmax approximated solution to the original problem in \eqref{eq:int_p0}.
\end{defi}

Suggested by the zero temperature limit in Proposition~\ref{prop:gumbel}, we know the objective function $\hat{f}_{\mathrm{soft}}^{N,\tau}$ in \eqref{eq:p1} converges to the empirical estimator in \eqref{eq:approx1} in probability as $\tau$ goes to $0^{+}$. While $\hat{f}_{\mathrm{soft}}^{N,\tau}$ is differentiable with respect to $\bm{\alpha}$, it is not identical to the empirical estimator in \eqref{eq:approx1} for non-zero temperature. In other words, there is a trade-off between small temperatures, where samples are close to one-hot but the standard deviation of the gradients is large,
and large temperatures, where samples are smooth but the standard deviation of the gradients is small. In practice, we shall tune the temperature $\tau$ from high to a small but non-zero value \cite{2016Categorical}.

\begin{remark}\label{rk:1}
Since the objective function of the Gumbel-Softmax reparameterization problem in \eqref{eq:p1} is random, the approximated solution in \eqref{eq:approx_min} should be viewed as a random variable as well. From the first glance, the randomness of $\hat{f}_{\mathrm{soft}}^{N,\tau}$ is inconsistent with the deterministic feature of the original integer programming problem in \eqref{eq:int_p0}. However, in practice, the uncertainty of the approximated solution in \eqref{eq:approx_min} is beneficial in several aspects. For example, since we are solving a series of unconstrained approximated problems based on the soft-loss function in \eqref{eq:soft_loss}, the feasibility of the solution in \eqref{eq:approx_min} is not guaranteed. Thus, the randomness allows us to generate samples of the approximated solution and select the one that produces the lowest objective function value among all the samples in the feasible domain.
\end{remark}

\begin{remark}\label{rk:2}
    We have to admit that our framework suffers from the curse of dimensionality. In the reparameterized problem \eqref{eq:p1}, the decision variable $\bm{\alpha}$ is of the same dimension as the one-hot vector that encodes the decision variable $y$ in the original integer programming problem \eqref{eq:int_p0}, which grows exponentially as the problem dimension increases.  {\color{black} In practice, the dimensionality issue makes it impossible to exactly solve the reparameterization problem \eqref{eq:p1}. However, the continuous feature of $\bm{\alpha}$ in \eqref{eq:p1} allows us to implement the neural networks. In Section~\ref{sec:network}, we will model the decision variable $\bm{\alpha}$ using neural networks in a decoupling manner.
    }
\end{remark}

\section{Problem Formulation}
\label{sec:problem}

\begin{figure}[ht!]
\center
\includegraphics[width=0.95\linewidth]{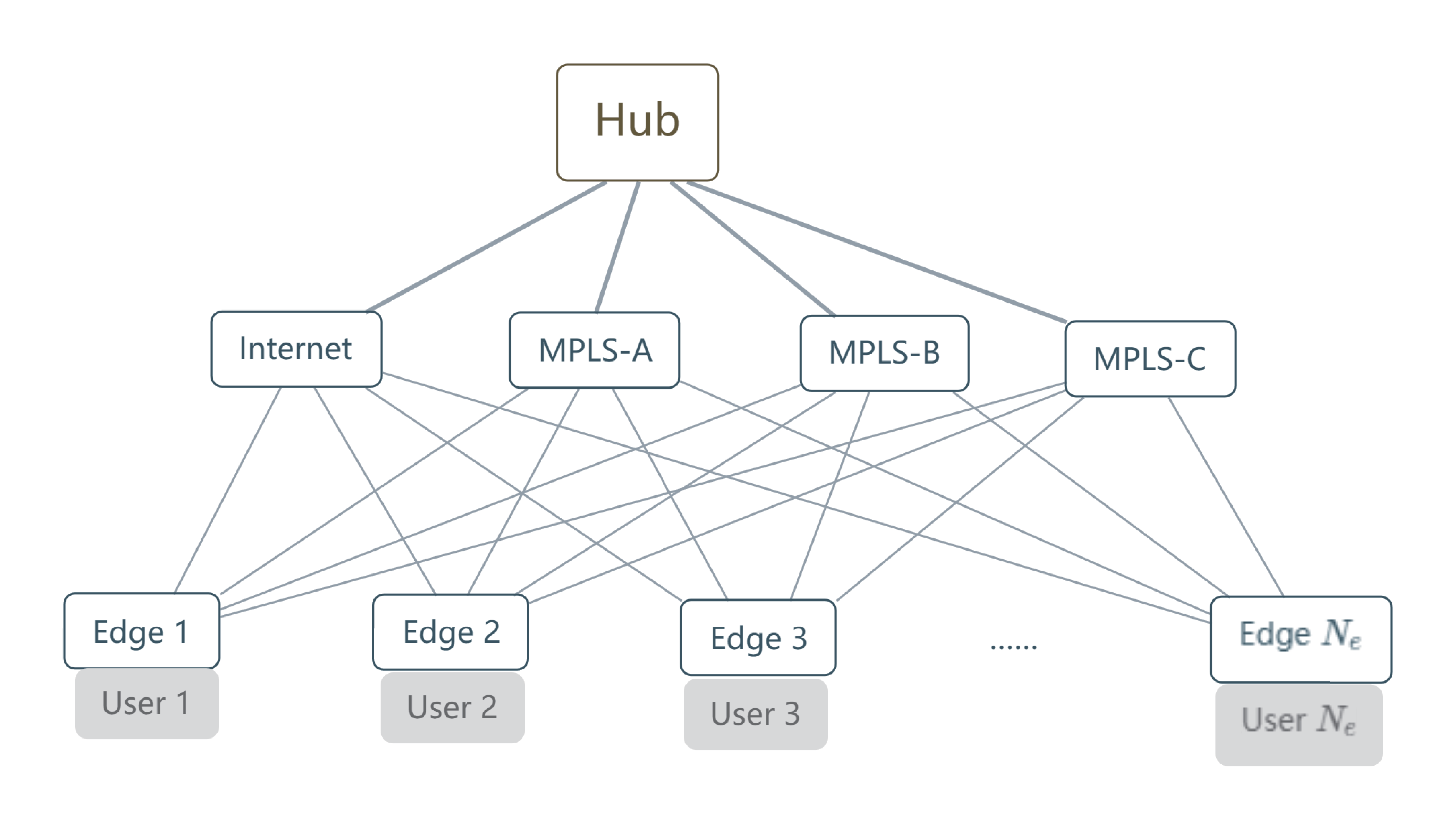}
\caption{{\color{black}The network topology of the SD-WAN.} The ``Internet'', ``MPLS-A'', ``MPLS-B'', and ``MPLS-C'' are four different ISPs. MPLS is the abbreviation of multi-protocol label switching, which enables ISPs to build intelligent networks that deliver various services over a single infrastructure \cite{2004Generalized}. 
  The four peering links between ISPs and Hub also contribute to the total cost based on their utilization.}
\label{fig:network}
\end{figure}

In this section, we formalize the offline version of the cloud network bandwidth costs model under a WAN of fixed topology in Figure~\ref{fig:network} containing a single hub and $N_{e}$ edges. Each edge represents a user in the SD-WAN (Figure~\ref{fig:SD-WAN}). These two concepts are equivalent in our discussion. Each edge has $K$ different traffic types and connects to $N_{I}$ Internet Service Providers (ISPs). For simplicity, we adopt a constant value of 8 for the parameter $K$ and 4 for the parameter $N_{I}$, but our method can be extended to arbitrary network size. Links in the topology are billed individually according to their percentile utilization. Utilization-based, per-megabit billing is the industry standard for paid peer and transits ISP contracts\cite{Singh21}. This billing model is also considered in our paper. To determine the billable bandwidth from the network utilization, ISPs measure the average utilization of peering links in five-minute intervals in both inbound and outbound directions denoted by $\{\bar{f}^{t}\}_{t=1}^{T}, \{\underline{f}^{t}\}_{t=1}^{T}$, respectively, where $T$ corresponds to the total number of five-minute intervals in a single billing cycle. For example, $T=8640$ for a monthly billing cycle with $30$ days. The billable bandwidth of the billing cycle is
\begin{equation}\label{eq:billable}
z = \max \{ g_{95}(\{\bar{f}^{t}\}_{t=1}^{T}), g_{95}(\{\underline{f}^{t}\}_{t=1}^{T}) \},  
\end{equation}
where $g_{95}$ denotes the $95^{\text{th}}$ percentile function, which is the same as the $k$-$\max$ function with $k = T/20$.

Let $E = \bigcup\limits_{n=1}^{N_{e}} E_{n}$ denote the set of edge lines, where $E_{n} = \{e_{n,1}, e_{n,2},e_{n,3}, e_{n,4}\}$ represents the four peering links physically connected to edge-$n$. We use $L = \{\ell_1, \ell_2, \ell_3, \ell_4\}$ to denote the four peering links between ISPs and Hub. Given the traffic demands within a billing cycle, we aim to minimize the sum of bandwidth costs on peering links $E$ and $L$. We need a series of concepts and notations to formulate the corresponding mathematical model.

\textbf{Decision variables.} In each five-minute interval, for edge-$n$,
the traffic allocation scheme assigns the network flow from each type of traffic demand to at least one peering link in $E_{n}$. Let $Y = \{ y_{e,k,n}^{t}\}$ be the set of 0-1 decision variables, where $y_{e,k,n}^{t} = 1$ means at time slot $t$, the type-$k$ traffic demand on edge-$n$ is assigned to peering link $e\in E_{n}$. The dimension of the decision variable $Y$ is $K*N_{I}*N_{e}*T = 32N_{e}T$.

\textbf{Objective function.} Let $z_{e}$, $e\in E$ and $z_{\ell_{i}}$, $\ell_{i}\in L$ be the billable bandwidth of all peering links defined by the $95^{\text{th}}$ percentile of the snapshots of the utilization as in \eqref{eq:billable}. The cost incurred on each link is the product of the peering link rate and the amount of the billable bandwidth that exceeds the basic link capacities, that is, let $\sigma(x) = \max \{x, 0\}$, and we have
\begin{equation}\label{eq:cost}
 f = \sum_{e\in E} r_{e} \sigma ( z_{e} - c_{e}^{b}) + \sum_{i=1}^{4}  r_{\ell_i} \sigma ( z_{\ell_i} - c_{\ell_i}^{b}),
\end{equation}
where $c_{e}^{b}$ and $c_{\ell_{i}}^{b}$ denote the basic link capacities of the link $e\in E$ and $\ell_{i}\in L$, respectively. In the actual usage, the customer commits to pay a fixed rent for each billing cycle to the ISP, which is not shown in the cost function in \eqref{eq:cost} since it does not affect the minimization problem.

\textbf{The split ratio of the traffic volume.} At each edge, when a traffic demand is assigned to more than one peering link, the demand is split, and the split ratio is proportional to the basic link capacities of the connected peering links. In particular, let $(\bar{d}_{k,n}^{t},  \underline{d}_{k,n}^{t})$ denote the the averaged type-$k$ inbound/outbound  traffic demand on edge $n$ at time slot $t$, then the inbound/outbound traffic carried by peering link $e\in E_{n}$ are given by
\begin{equation}\label{eq:flow_x}
\bar{x}_{e,k,n}^{t} = \frac{y_{e,k,n}^{t}c_{e}^{b}}{\sum\limits_{e\in E_{n}} y_{e,k,n}^{t}c_{e}^{b}}\cdot \bar{d}_{k,n}^{t}, \; \underline{x}_{e,k,n}^{t} = \frac{y_{e,k,n}^{t}c_{e}^{b}}{\sum\limits_{e\in E_{n}} y_{e,k,n}^{t}c_{e}^{b}}\cdot \underline{d}_{k,n}^{t}.
\end{equation}
The splitting in \eqref{eq:flow_x} naturally satisfies the 
traffic demand constraints, that is,
\begin{equation*}
\bar{d}_{k,n}^{t} = \sum_{e\in E_{n}} \bar{x}_{e,k,n}^{t}, \quad \underline{d}_{k,n}^{t} = \sum_{e\in E_{n}} \underline{x}_{e,k,n}^{t}.
\end{equation*}
Here, the average traffic demand $(\bar{d}_{k,n}^{t},  \underline{d}_{k,n}^{t})$ and the link capacities $c_{e}^{b}$ are part of the problem parameters (i.e., $\theta$ in \eqref{eq:int_p0}), which are available under the offline setup. 

We sum $\bar{x}_{e,k,n}^{t}$ and $\underline{x}_{e,k,n}^{t}$ over different types of traffic demands, and obtain the inbound/outbound traffic on the edge links
\begin{equation*}
\bar{f}^{t}_{e,n} = \sum_{k=1}^{8}  \bar{x}^{t}_{e, k,n}, \quad \underline{f}_{e,n}^{t} =  \sum_{k=1}^{8}  \underline{x}^{t}_{e, k,n}, \quad \forall e\in E_{n},
\end{equation*}
which induce the billable bandwidth $z_e$ based on \eqref{eq:billable}, i.e.,
\begin{equation*}
z_{e} = \max \{g_{95}(\{\bar{f}^{t}_{e,n}\}_{t=1}^{T}), g_{95}(\{\underline{f}^{t}_{e,n}\}_{t=1}^{T})\}, \quad e\in E_{n},
\end{equation*}
where $n = 1,2,\dots, N_{e}$. Correspondingly, for the traffic on the ISP links, we sum $\bar{f}^{t}_{e,n}$ and $\underline{f}^{t}_{e,n}$ over $n$, respectively, which lead to the billable bandwidth $z_{\ell_i}$, $i=1,2,3,4$. The complete problem formulation is presented in \eqref{eq:milp_nonlin}.

\textbf{Constraints.} From the previous discussion, we have seen that the traffic allocations are subject to constraints on link capacities and traffic demand. Besides, the service level agreement (SLA), e.g., \cite{2014Service}, introduces another set of constraints, which identify whether a peering link satisfies the requirement of the traffic demand. In reality, since the environment of the WAN is not static, such constraints change over time. In the model, we simplify the SLA constraints into the concept of admissible peering link, denoted by $E_{k,n}$. As a subsect of $E_{n}$, the type-$k$ demand on edge-$n$ can only be assigned to links in $E_{k,n}$. Under the assumption, $E_{k,n}$ is part of the problem data given in advance. Express in terms of equality constraints, and we have, for any $k$ and $n$, $y_{e,k,n}^{t} = 0$, $\forall e\not \in E_{k,n}$, which hold for all time slots $t$. 

For the link capacities, besides the basic link capacities that appeared in the objective function \eqref{eq:cost} and the split ratio \eqref{eq:flow_x}, we shall also introduce the upper bounds for the billable bandwidth and the link utilization, denoted by $c_{e}^{m}$/$c_{\ell_i}^{m}$ (maximum link capacity) and $c_{e}^{M}$/$c_{\ell_i}^{M}$ (physical maximum link capacity), respectively. Unlike the physical maximum link capacity determined by the infrastructure, the ISP chooses the maximum link capacity to set the upper bound for the billing rate $z_{e}$ and $z_{\ell_{i}}$ in the cost \eqref{eq:cost}, i.e.,
\begin{equation*}
    z_{e} \leq c_{e}^{m}, \; \forall e\in E, \quad z_{\ell_{i}} \leq c_{\ell_{i}}^{m}, \; i=1,2,3,4.
\end{equation*}

\begin{table}[ht!]
\def\arraystretch{1.5}
\centering
\caption{Classification of the parameters (problem data) in \eqref{eq:milp_nonlin}.}
\begin{tabular}{ l c}
 \hline
 Static parameters & $r_{e}, r_{\ell_i}, E_{k,n}, c_{e}^{b},c_{e}^{m}, c_{e}^{M},c_{\ell_{i}}^{b},c_{\ell_{i}}^{m}, c_{\ell_{i}}^{M}$  \\ \hline
 Dynamic parameters & $\bar{d}_{k,n}^{t},  \underline{d}_{k,n}^{t}$ \\ \hline
\end{tabular}
\smallskip
\label{tab:2}
\end{table}

\begin{table}[ht!]
\def\arraystretch{1.5}
\centering
\caption{List of variables and parameters in the integer programming problem \eqref{eq:milp_nonlin}.}
\begin{tabular}{ c l} 
\hline
{\color{black} $N_{e}$} & {\color{black} number of edges/users} \\ \hline
{\color{black} $T$} & {\color{black} number of intervals in a billing cycle }\\ \hline
 $Y = \{y_{e,k,n}^{t}\}$ & decision variable\\ \hline
 $r_{e}, r_{\ell_{i}}$ & peering link rate \\ \hline
  $z_{e}, z_{\ell_{i}}$& billable bandwidth \\ \hline
 $\bar{d}_{k,n}^{t},  \underline{d}_{k,n}^{t}$ & inbound/outbound demand  \\ \hline
 $\bar{x}_{e,k,n}^{t}, \underline{x}_{e,k,n}^{t}$ & splitting of inbound/outbound demand
 \\ \hline
$\bar{f}^{t}_{e,n}, \underline{f}^{t}_{e,n}$   &inbound/outbound traffic on edge link  \\  \hline
$\bar{X}_{\ell_{i}}^{t}, \underline{X}_{\ell_{i}}^{t}$ &  inbound/outbound traffic on ISP link \\ \hline
$E_{n}$ &  set of peering link\\ \hline
$E_{k,n}$ &  set of admissible peering link \\ \hline
$c_{e}^{b}, c_{e}^{m}, c_{e}^{M}$ &  edge link capacities \\ \hline
$c_{\ell_{i}}^{b},c_{\ell_{i}}^{m}, c_{\ell_{i}}^{M}$ & ISP link capacities \\
\hline
\end{tabular}
\smallskip
\label{tab:1}
\end{table}

To summarize our discussion, we introduce the following integer programming problem, which models the network scheduling problem for edge-cloud \textcolor{black}{networking}.
\begin{equation}\label{eq:milp_nonlin}
\begin{split}
& \min_{Y} \quad f \quad \text{s.t.} \\
&\sum_{e\in E_{k,n}} y_{e,k,n}^{t} \geq 1, \quad \forall e\in E_{n}, \; \forall n,t; \\
& \bar{f}^{t}_{e,n} = \sum_{k=1}^{8}  \bar{x}^{t}_{e, k,n}, \quad \underline{f}_{e,n}^{t} =  \sum_{k=1}^{8}  \underline{x}^{t}_{e, k,n},  \\
&\bar{f}^{t}_{e,n}, \; \underline{f}_{e,n}^{t} \leq c_{e}^{M}, \quad \forall e\in E_{n}, \; \forall n,t;\\
& \bar{X}_{\ell_{i}}^{t} = \sum_{n=1}^{N_{e}} \bar{f}^{t}_{e_{n,i},n}, \quad \underline{X}_{\ell_{i}}^{t} = \sum_{n=1}^{N_{e}}  \underline{f}_{e_{n,i},n}^{t}, \\
&\bar{X}_{\ell_{i}}^{t}, \; \underline{X}_{\ell_{i}}^{t} \leq c_{\ell_{i}}^{M}, \quad i = 1,2,3,4; \\
&z_{e} = \max\left\{g_{95}\big(\{\bar{f}^{t}_{e,n}\}_{t=1}^{T}\big),\; g_{95}\big(\{\underline{f}_{e,n}^{t}\}_{t=1}^{T} \big) \right\}, \\
&z_{\ell_{i}} =   \max\left\{g_{95}\big(\{\bar{X}_{\ell_{i}}^{t}\}_{t=1}^{T}\big),\; g_{95}\big(\{\underline{X}_{\ell_{i}}^{t}\}_{t=1}^{T} \big) \right\}, \\
& z_{e} \leq c_{e}^{m} , \quad \forall e \in E_{n}, \quad n = 1,2,\dots, N_e;  \\
&  z_{\ell_i} \leq c_{\ell_i}^{m}, \quad i = 1,2,3,4,
\end{split}
\end{equation}
where $\bar{x}_{e,k,n}^{t}, \underline{x}_{e,k,n}^{t}$ are given by \eqref{eq:flow_x} and the objective function $f$ is defined in \eqref{eq:cost}. The variables in \eqref{eq:milp_nonlin} are summarized in Table~\ref{tab:1}. We want to emphasize that although the objective function $f$ in \eqref{eq:cost} is the sum of the cost on each link, the problem \eqref{eq:milp_nonlin} cannot be decomposed into low-dimensional sub-problems since the billable bandwidth of the ISP link $z_{\ell_{i}}$ non-linearly depends on the entire edge traffic. Nevertheless, we can linearize the problem \eqref{eq:milp_nonlin} by adding intermediate variables to the system. (See Appendix~\ref{app:linear} for the details) Although we stick to the form in \eqref{eq:milp_nonlin} in the later discussion, the linear representation is useful for conventional methods.
\begin{remark}\label{rk:3}
Although problem \eqref{eq:milp_nonlin} corresponds to the offline version of the network scheduling model in the sense that the problem parameters are given in advance\cite{NetworkCheng}, we should think of \eqref{eq:milp_nonlin} as a family of integer programming problems of the form in \eqref{eq:int_p0} parameterized by the problem data. We classify the problem parameters into static and dynamic parameters as in Table~\ref{tab:2}. Here, the static parameters, e.g., the link capacities, are physically determined once the network's topology (Figure~\ref{fig:network}) is chosen. In comparison, the traffic demands, decided by the users, are random and constantly changing over different billing cycles. In other words, we interpret the static parameter as global parameters among the family of problems, while the dynamic parameters distinguish the problems within the family. Later in Section~\ref{sec:network} and Appendix~\ref{app:DataGeneration}, we follow such interpretations and generate test problems by randomly sampling the dynamic parameters subject to fixing static parameter values.
\end{remark}

Another perceptive insight is that link utilization during $5\%$ of time slots does not contribute to the cost under the $95^{th}$-percentile billing policy. This means that the top $5\%$ traffic in any billing month is free if it does not exceed the physical maximum link capacity. Similar to the works like \cite{Singh21}, instead of searching for the best traffic allocation scheme precisely, i.e., the global minimizer of \eqref{eq:milp_nonlin}, our goal is finding an ``end-to-end'' map between the program data (Table~\ref{tab:2}) and the allocation scheme. Such a map should be capable of efficiently generating allocation schemes of reasonable quantity for problems in the family.

\section{Neural Network Architecture and Algorithm} \label{sec:network}

In this section, we introduce the Gumbel-Softmax Sampling Network (GSSN), which is designed for solving the integer programming problem \eqref{eq:milp_nonlin}. In the following sections, we sequentially discuss the data preprocessing, neural network architecture, training, and implementation of the GSSN.

{\color{black}
The key idea of GSSN is modeling a probability mass vector over the space of admissible schemes by neural networks. As we have pointed out in Remark~\ref{rk:2}, this framework suffers from the curse of dimensionality. Regarding the dimension of the decision variable in the target problem  \eqref{eq:milp_nonlin}, in general, it is not applicable to consider the entire allocation schemes of all edges. For the sake of computational efficiency and real-world applications, we process different time slots, types of traffic, and users separately in a decoupled manner. As a result, the GSSN architecture and training algorithm are designed for sampling candidate schemes for the type-$k$ traffic of user-$n$ at time step $t$.}

In the discussion, we use the ``$\mathrm{Unfold}$'' and ``$\mathrm{Reshape}$'' operations to convert matrix-valued data to vector-valued data and vice versa, respectively. Recall that we say a column vector $B\in \mathbb{R}^{m\times n}$ is an unfolding of a matrix $A\in \mathbb{R}^{m\times n}$, denoted by $B = \mathrm{Unfold}(A)$, if $B$ consists of all the column vectors of $A$ in row order. Correspondingly, the inverse operation, reshaping the $m \times n$ dimensional column vector $B$ into a matrix $A$ of size $(m,n)$, is denoted as
$A = \mathrm{Reshape}(B,n)$.

\subsection{The Neural Network Input} \label{sec:input}

This subsection provides a comprehensive account of data preprocessing for the GSSN. The GSSN aims to map the local information of inbound and outbound traffic, $(\bar{d}^t_{k,n},\underline{d}^t_{k,n})$, to a probability distribution on the space of traffic allocation schemes. To achieve this objective, we process all types of inbound and outbound traffic for all edges/users simultaneously in parallel. To illustrate, we consider the $k^{\text{th}}$ type of traffic $(\bar{d}^t_{k,n},\underline{d}^t_{k,n})$, of user $n$ at time $t$ as the target and assume their set of admissible peering links as $E_{k,n} = \{e_1, e_2\}$. In the subsequent sections, we expound on the formation of input matrices $I^t_{k,n}$ that corresponds to $(\bar{d}^t_{k,n},\underline{d}^t_{k,n})$. In cases where traffic demand is not explicitly distinguished as inbound or outbound, the symbol ${d}^t_{k,n}$ may denote traffic demand. This symbol may be substituted with $\bar{d}^t_{k,n}$ or $\underline{d}^t_{k,n}$ as appropriate. 
Every possible way of selecting $\{y_{e_j,k,n}^t\}_{j = 1,2,3,4}$ corresponds one-to-one with the non-empty subsets of $E_{k,n} = \{e_1, e_2\}$. Therefore, there are at most three available allocation schemes for the demand ${d}^t_{k,n}$. To capture the potential values of $\{y_{e_j,k,n}^t\}_{j = 1,2,3,4}$, we use the matrix $YP_{d_{k,n}^t}$, where $YP_{d_{k,n}^t}[i,j]$ denotes the value of $y_{e_j,k,n}^t$ for the $j^{th}$ edge in the $i^{th}$ scheme.

\begin{equation}\label{eq: yMatrix}
    YP_{d_{k,n}^t} = \begin{bmatrix}
        1&0&0&0\\
        0&1&0&0\\
        1&1&0&0\\
        0&0&0&0\\
        \vdots&\vdots&\vdots&\vdots\\
    \end{bmatrix} \in \{0,1\}^{P\times 4}.
\end{equation}
Here, $P= 2^{N_{I}}-1= 15$, corresponds to the number of nonempty subsets of set $E_{n}$. The first three rows of $YP_{d_{k,n}^t}$ in \eqref{eq: yMatrix} represent the three possible allocation schemes, while the elements in rows 4 to 15 are all zero. The purpose of this padding operation is to unify the input matrices of different traffic demands for different users. In general, if there are $s$ optional edges in $E_{k,n}$, then the first $2^s-1$ rows of the corresponding $YP_{d_{k,n}^t}$ matrix represent the possible allocation schemes for admissible peering links, while the remaining rows are filled with zeros.

In accordance with the split ratio of traffic volume definition provided in Section~\ref{sec:problem}, we can represent the traffic split ratio matrix $W_{d_{k,n}}$, which corresponds to $YP_{d_{k,n}^t}$, as follows. Specifically, the element $W_{d_{k,n}}[i,j]$ indicates the split ratio of traffic volume that should be adhered to on the $j^{th}$ edge for the $i^{th}$ allocation scheme of traffic demand $d_{k,n,t}$.
\begin{equation}\label{eq: TraffiSplitMatrix}
    W_{d_{k,n}} = \begin{bmatrix}
        1&0&0&0\\
        0&1&0&0\\
        \frac{c_{e_1}^b}{\sum_{i=1,2}c_{e_i}^b}&\frac{c_{e_2}^b}{\sum_{i=1,2}c_{e_i}^b}&0&0\\
        0&0&0&0\\
        \vdots&\vdots&\vdots&\vdots\\
    \end{bmatrix}\in \mathbb{R}^{15\times 4}
\end{equation}
It is worth noting that every element in the traffic split ratio matrix $W_{d_{k,n}}$ is derived from the static parameter $c_e^b$. Consequently, $W_{d_{k,n}}$ is a quantity that remains invariant with respect to variations in time $t$ and demand $d_{k,n,t}$. This also explains why the subscript of  $W_{d_{k,n}}$ does not contain the parameter $t$.

To obtain the allocation of inbound and outbound traffic $(\bar{d}_{k,n}^{t},  \underline{d}_{k,n}^{t})$ on the selected peering links under the selection scheme, we can perform the following calculations:
  \begin{equation}
     G_{\bar{d}_{k,n}^{t}} = \bar{d}_{k,n}^{t} * W_{d_{k,n}};\quad G_{\underline{d}_{k,n}^{t}} = \underline{d}_{k,n}^{t} * W_{d_{k,n}}.
  \end{equation}
$G_{d^t_{k,n}}[i,j]$ represents the amount of traffic allocated to the $j^{th}$ edge in the $i^{th}$ allocation scheme for the traffic demand $d^t_{k,n}$.

The final input matrix $I_{k,n,t}$ can be obtained through the following processing.
\begin{equation}\label{eq:iknt}
    \begin{split}
    &\bar{G}_{k,n,t} = \mathrm{Unfold}(G_{\bar{d}_{k,n}^{t}}), \;
    \underline{G}_{{k,n,t}} = \mathrm{Unfold}(G_{\underline{d}_{k,n}^{t}})\\
    &I_{k,n,t} = 
    \begin{bmatrix}
    \bar{G}_{k,n,t}[1]&\underline{G}_{{k,n,t}}[1]& c_{e_1}^b&c_{e_1}^m\\[6pt]
    \bar{G}_{k,n,t}[2]&\underline{G}_{{k,n,t}}[2]& c_{e_2}^b&c_{e_2}^m\\[6pt]
    \bar{G}_{k,n,t}[3]&\underline{G}_{{k,n,t}}[3]& c_{e_3}^b&c_{e_3}^m\\[6pt]
    \bar{G}_{k,n,t}[4]&\underline{G}_{{k,n,t}}[4]& c_{e_4}^b&c_{e_4}^m \\[6pt]
    \dots &\dots &\dots&\dots\\[6pt]
    \bar{G}_{k,n,t}[60]&\underline{G}_{{k,n,t}}[60]& c_{e_4}^b&c_{e_4}^m \\[6pt]
    \end{bmatrix}\in \mathbb{R}^{60 \times 4}
    \end{split}
\end{equation}

The row index $i$ of $I_{k,n,t}$ \eqref{eq:iknt} corresponds to the $j^{th}$ edge in the $p^{th}$ scheme, satisfying the following equation:
  \begin{equation*}
  i = 4*p + j,\quad p\in \{0,1,\dots, 14\},\; j\in\{1,2,3,4\}.
  \end{equation*}
Specifically, $I_{k,n,t}[i,1]$ and $I_{k,n,t}[i,2]$ denote the allocation of inbound/outbound traffic demand of flow ${d}_{k,n}^{t}$ on the $j^{th}$ edge in the $p^{th}$ allocation scheme. On the other hand, $I_{k,n,t}[i,3]$ and $I_{k,n,t}[i,4]$ represent the capacity information of the $j^{th}$ edge. The final input matrix $I$ for problem \eqref{eq:milp_nonlin} is obtained by concatenating the sub-matrices of $I_{k,n,t}$ in ascending order of traffic type $k$, user $n$, and time slot $t$.

\subsection{Neural Network Architecture}

\label{sec:NN_Arc}

{\color{black}
The GSSN architecture comprises three encoders: the link encoder, the program encoder, and the ranking autoencoder. The different types of traffic from the users at each time period are treated in a decoupled manner. For example, as a crucial part of the GSSN, the ranking autoencoder only takes the feature information of $P$ candidate schemes for user $n$'s $k^{\text{th}}$ type of traffic at time step $t$ and outputs the selection probabilities for the $P$ candidate schemes with respect to the input traffic. For input data preprocessing, the link encoder and program encoder serve to compress and extract features from different schemes. The link encoder aims to encode multiple feature information, such as the traffic demands and link capacities, associated with each link into a one-dimensional representation, while the program encoder compresses the feature information for the four edges of each candidate scheme. 

After numerically investigating several conventional activation functions, we consider $ReLU6(x)\triangleq \min\{\max\{0,x\},6\}$ as the ideal activation function for the implementation of GSSN. In the PyTorch framework, $ReLU6(x)$ is a widely-used preset activation function \cite{ReLU6}.
 
 }

\textbf{The Link Encoder:} The link encoder comprises a fully connected, feed-forward Neural Network(FNN) with three hidden layers and 8 neurons. Upon input, data $I$ is processed by a link encoder, yielding a column vector $S$ of dimensions $T*N_{e}*K*P$. Each row of $S$ conveys the compressed encoding characteristics of the $k^{th}$ traffic type for user $n$ at time $t$ on each peering link under scheme $p$. Define the matrix $S^{\prime} = \mathrm{Reshape}(S,4)$. Each row of $S^{\prime}$ represents the encoding information of the scheme $p$ selected by the $k^{th}$ traffic of user $n$ at time $t$.

\textbf{The Program Encoder:} The program encoder employs a fully connected neural network architecture identical to the link encoder. The input matrix, denoted as $S$, undergoes encoding by the program encoder, compressing each 4-dimensional feature vector into a one-dimensional scalar. The output is a matrix $V$ with dimensions $(T*N_{e}*K*P,1)$. Subsequently, $V$ is reshaped into $V^{\prime}= \mathrm{Reshape}(V)$.

\textbf{The Ranking AutoEncoder:} The Ranking AutoEncoder constitutes a prototypical autoencoder architecture comprising an encoder and a decoder. The components of these two neural network structures exhibit symmetry, encompassing $6$ layers containing $58$ neurons. The Ranking AutoEncoder operates to map all schemes of the $k^{\text{th}}$ type of traffic of user n at time t to a specific probability distribution function. This mapping enables subsequent Gumbel-Softmax sampling. After inputting $V^{\prime}$ into the Ranking AutoEncoder, the output is a certain probability distribution $\alpha$. $\alpha$ is a matrix with dimensions $(T*N_{e}*K, P)$, where each row represents the probability distribution of all schemes of the $k^{th}$ type of user traffic $n$ at time $t$.

\textbf{Masked Gumbel-Softmax Sampling:} Upon encoding via three preceding encoders, the probability distribution $\alpha$ of the scheme is obtained. It should be noted that invalid schemes are also inputted to maintain constant dimensions of the encoder’s input matrix. To prevent GSSN from outputting invalid schemes, $\alpha$ is multiplied by a mask of identical dimensions. The mask assigns a probability of 0 to elements in $\alpha$ corresponding to invalid schemes. As discussed in  Section \ref{sec:softmax}, the range of $\alpha$ values is $[0,\infty)$. Following the mask’s action, $\alpha$ is obtained. Gumbel-Softmax performs a maximum value operation to select the index of the One-Hot vector element equal to 1, precluding the selection of invalid schemes. For the hyperparameters $\tau$ of the Gumbel-Softmax in Section~\ref{sec:softmax}, we adopted a strategy of linearly decreasing monotonically as the epoch increases.

At this point, we obtain the selection scheme for the inbound and outbound transmission lines for different traffic of user $n$ at time $t$. We can substitute equation \eqref{eq:milp_nonlin} to calculate the loss function and perform gradient training.

\subsection{How to Use GSSN} 

\begin{figure}[ht!]
\centering
\begin{adjustwidth}{-0.1in}{-0.2in}
  \includegraphics[width=.95\linewidth]{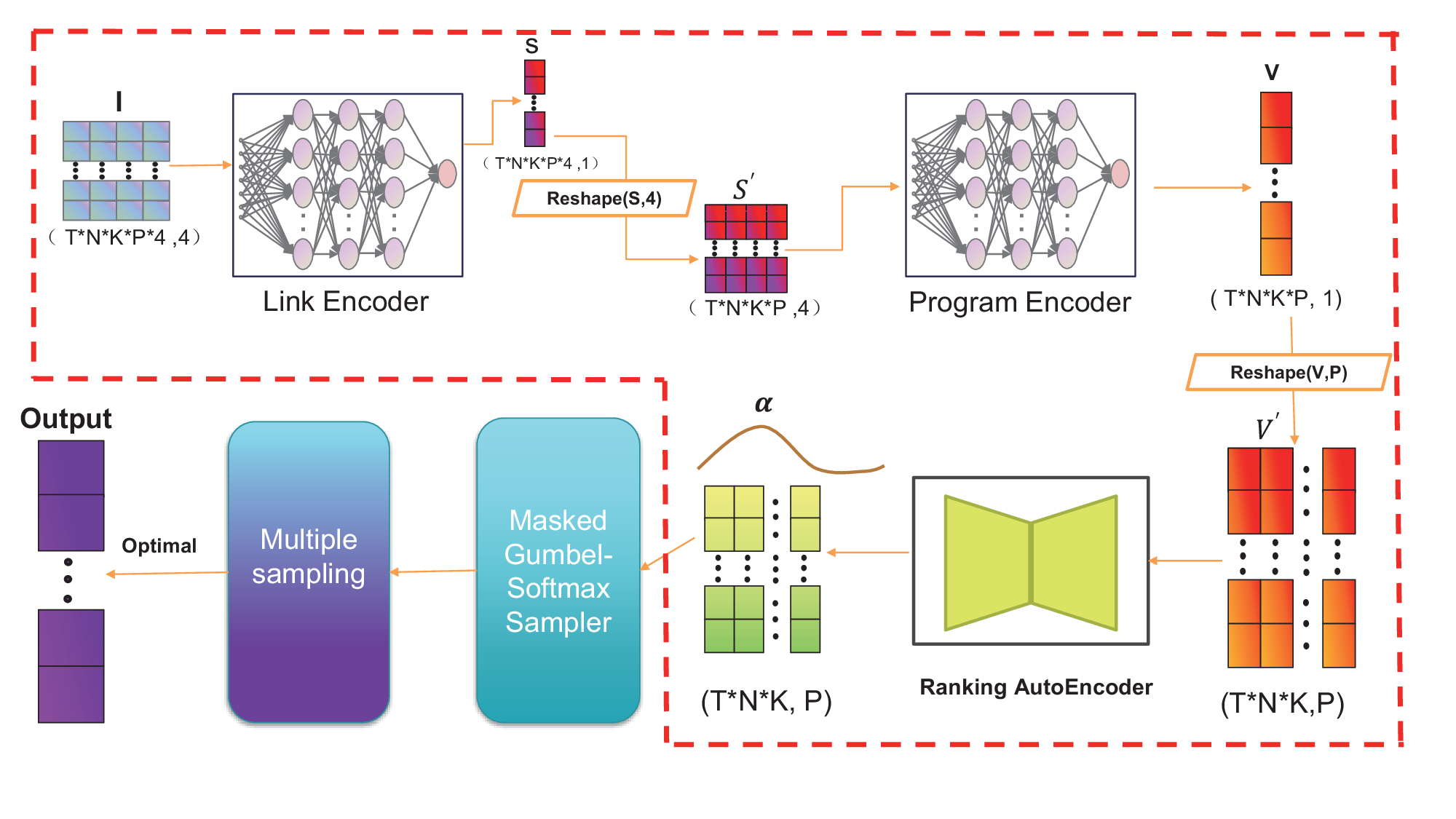}
  \end{adjustwidth}
  \caption{The schematic of the network. The transformation context of input data between three encoders - the link encoder, the program encoder, and the Ranking AutoEncoder - is represented by the red dashed line box. The outputs of GSSN are obtained through multiple sampling of Gumbel-Softmax to derive the final solution probability distribution.}
\label{fig:NN}
\end{figure}

The GSSN network’s training process is outlined in Algorithm~\ref{alg:TrainingGSSN}. After training the GSSN, the information matrix $I$ of the test set problem can be input. By repeating the input of matrix  $N_{sampling}$ times, the GSSN samples $N_{sampling}$ different scheme selections. We choose the feasible scheme with the smallest objective function value from the schemes as the final output scheme of the GSSN algorithm. The complete GSSN network architecture can be seen in Figure~\ref{fig:NN}.

\begin{algorithm}[ht!]
\caption{Training Gumbel-Softmax Sampling Network}\label{alg:TrainingGSSN}
\begin{algorithmic}
\STATE 
\STATE {\textbf{Input: }} Generate input matrixs $\{I_n\}_{n=1}^{N_{data}}$ of $N_{data}$ problems \eqref{eq:milp_nonlin} with $N_e$ users, $T$ time periods, $K$ types of traffic, and a maximum of $P$ selection schemes for each traffic.
\STATE {\textbf{Input: }} $\theta$, initial GSSN parameters.
\STATE {\textbf{Output: }} $\hat{\theta}$, the trained parameters.
\STATE {\textbf{Hyperparameters: }} $N_{epochs} \in \mathbb{N}$, $\eta \in (0,\infty),$ $\tau_{start},\tau_{end} $ $ \in \mathbb{R}$
\STATE {\textbf{for}} $i=1,2,\dots,N_{epochs}$ do
 \STATE \hspace{0.5cm}  $\tau = \tau_{start} - \frac{i}{N_{epochs}}(\tau_{start}-\tau_{end})$
\STATE \hspace{0.5cm} {\textbf{for}} $n=1,2,\dots,N_{data}$ do
\STATE \hspace{1cm}  $S = LinkEncoder(I_{n}|\theta)$
\STATE \hspace{1cm}  $S^{\prime} = Reshape(S,4)$
\STATE \hspace{1cm}  $V = ProgramEncoder(S^{\prime},4|\theta)$
\STATE \hspace{1cm}  $V^{\prime} = Reshape(V,P)$
\STATE \hspace{1cm}  $\alpha(\theta) = RankingAutoEncoder(V^{\prime}|\theta)$
\STATE \hspace{1cm}  $\{y_{e,k,n}^t\}_{\forall e\in E_{k,n}, \forall n,t} \Longleftarrow  Sampling(\alpha(\theta)|\tau)$
\STATE \hspace{1cm} Compute $loss(\theta)$ by equation \eqref{eq:soft_loss}
\STATE \hspace{1cm} $\theta  = \theta -\eta\cdot \nabla loss(\theta)$
\STATE \hspace{0.5cm} {\textbf{end}}
\STATE {\textbf{end}}
\STATE \textbf{return}  $\hat{\theta} = \theta$
\end{algorithmic}
\label{alg1}
\end{algorithm}

We set the initial learning rate of our neural network training to 1e-4, and the parameter updates are performed using the Adam algorithm\cite{2014Adam}. As for the update of the hyperparameter $\tau$ in Gumbel-Softmax sampling, we adopt a linear annealing method based on the reference literature\cite{2016Categorical}. 

{\color{black}
To summarize, we developed GSSN to solve the offline scheduling problem, it can efficiently produce feasible solutions of superior quality. In the following section, we will demonstrate this point through numerical simulations.
}

\section{Numerical Results}
\label{sec:num}

This section provides numerical results from GSSN. The results include comparisons between the GSSN sampling method and random sampling, performance comparisons between GSSN warmup and Gurobi solving, and the generalization performance of GSSN.

\subsection{Problem Settings}\label{sec:setting}

All experiments are conducted on an Inter Core i7 laptop with 32GB RAM and accelerated using a 3070-8G GPU. We generate $200$ problems with $N_e = 10$ edges and $T = 48$ total time slots. In this study, $100$ problems are randomly selected as the training set, while the remaining $100$ are designated as the test set. \textcolor{black}{The training process for the model takes less than $15$ minutes.}

Initially, we independently generate all static parameters using uniform distributions. Subsequently, we independently generate dynamic parameters traffic demands for different problems based on a fixed network topology determined by the aforementioned static parameters. To control the problem difficulty, we establish a certain constraint relationship between the sum of all types of traffic demands for the user $n$ at time $t$ and the sum of the edge capacities as follows.
{\color{black}
\begin{equation} \label{eq:assumpDemand}
    \sum_{k=1}^{K} {d}^{t}_{k,n} \leq 2*\sum_{e\in E_{n}} c^b_{e} \leq \sum_{e\in E_{n}} c^m_{e}.
\end{equation}}
Due to the existence of traffic splitting ratios, Eq.~\eqref{eq:assumpDemand} cannot guarantee that a randomly selected solution satisfies all constraints of problem~\eqref{eq:milp_nonlin}, as demonstrated in later experiments with random networks. {\color{black}Further details regarding the experimental setup, including the network topology and sampling of the problem parameters, can be found in Appendix~\ref{app:DataGeneration}.}

We employed the following two comparative methods to evaluate the quality of the warmup generated by the GSSN sampler for problem \eqref{eq:milp_nonlin}.

\begin{itemize}
    \item \textbf{Gurobi}: a commercial software that solves mixed-integer programming problems. The academic version 9.1.2 is used in this study. Gurobi solves problem 
    \eqref{eq:milp_nonlin} using the linearization model developed in Appendix~\ref{app:linear}.
    \item \textbf{Random Sampling Network(RSN)}: the structure of the network is analogous to that of the GSSN, except for the users' inbound and outbound traffic link selection, which no longer adheres to the probability distribution obtained through neural network learning. Instead, a candidate solution is randomly generated from the set of admissible peering links $E_{k,n}$ based on the uniform distribution over its set of all possible schemes.
\end{itemize}

\subsection{Neural Network Training and the Sampling Distribution}

\begin{figure}[ht!]
\center
  \includegraphics[width=0.95\linewidth]{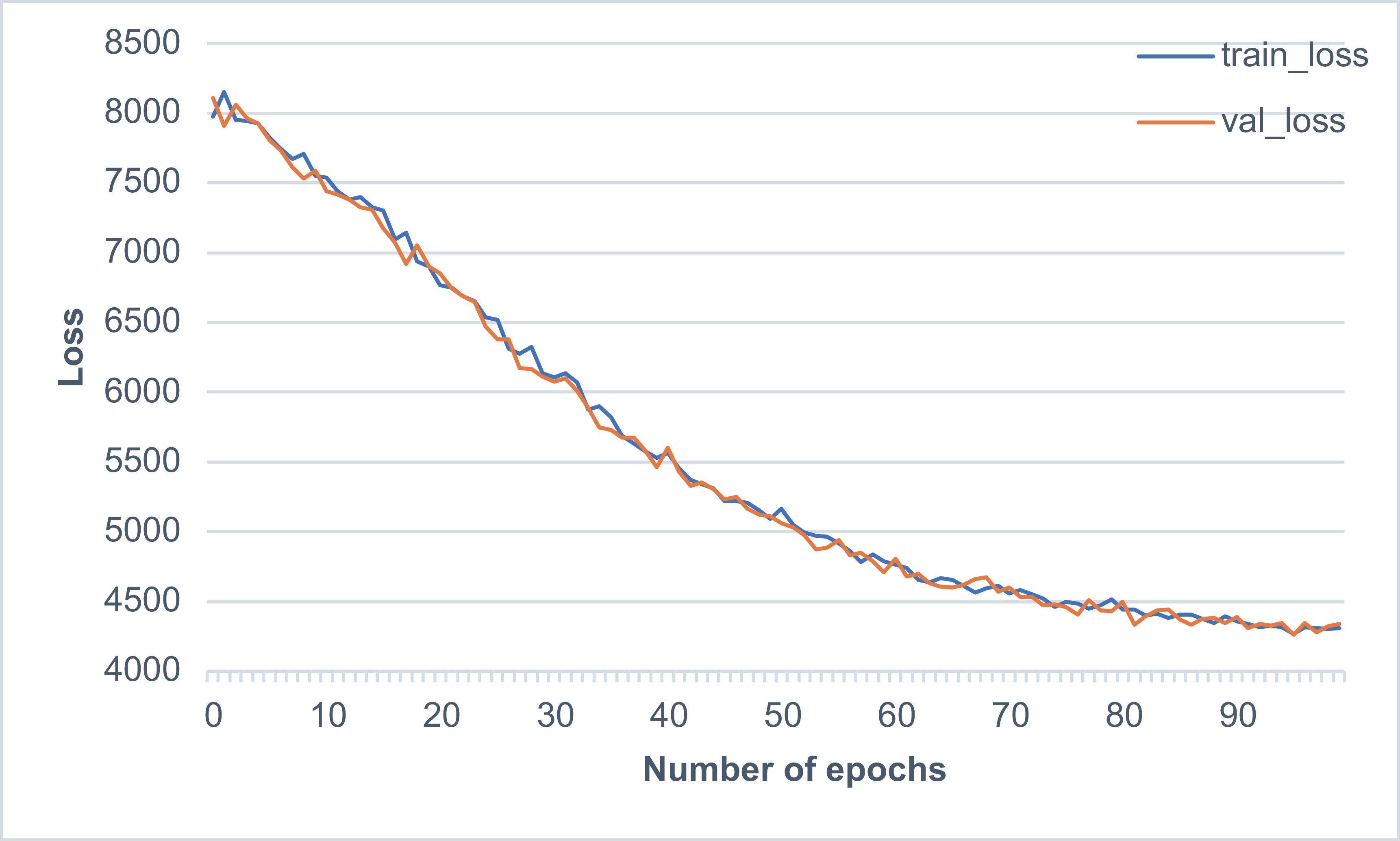}
  \caption{The averaged value of the soft-loss function of the training (blue curve) and testing problems (yellow curve). The soft-loss converges after 60 epochs. The training and testing errors behave similarly mainly because their problem data are sampled from the same distribution.}
\label{fig:loss}
\end{figure}

Training the neural network involved $100$ epochs with a batch size of 1. We take the average of the soft-loss functions over the test problems as the loss function in training. The decays of the loss function value for training and testing problems are similar, as suggested by Figure~\ref{fig:loss}. We observed no violation of the constraints during the training process, i.e., all GSSN samples are feasible solutions due to the mild difficulty of the problem controlled by the assumption \eqref{eq:assumpDemand}.

{\color{black}
It is worth mentioning that, due to the decoupled structure of GSSN (Section~\ref{sec:NN_Arc}), the number of parameters in GSSN to be trained is independent of the size of users($N_{e}$) and time intervals in a billing cycle($T$). Furthermore, note that the number of terms in the objective function $f$ in equation \eqref{eq:cost} increases linearly with respect to both $N_e$ and $T$. As a result, if we fix the size of the test problems, there exists an upper bound on the computational cost of the neural network training involved in GSSN, which linearly depends on $N_e$ and $T$, i.e., the training time of GSSN grows linearly with the size of users or time intervals in a billing cycle, under the same hardware environment.

}

\begin{figure}[ht!]
\center
  \includegraphics[width=1\linewidth]{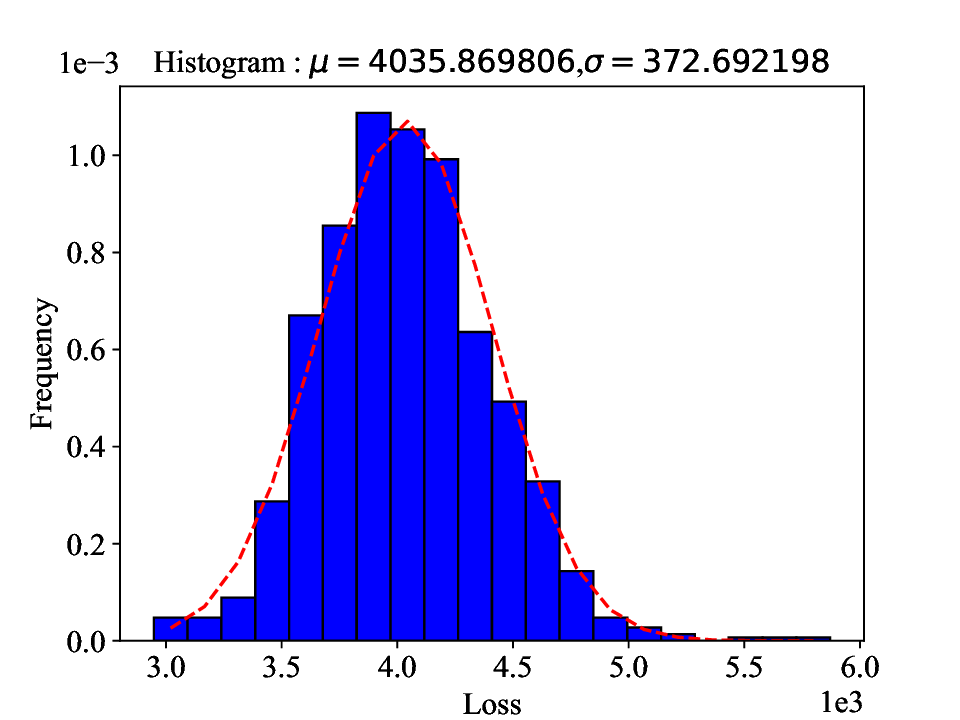}
  \caption{The histogram of the loss of the GSSN output. {\color{black}(We normalized the frequency so that the area of the bin corresponds to the probability, as do the rest of the histograms in the paper)} The output is random since we implement the Gumbel-Softmax trick in sampling the traffic allocations. We compute the value of the objective function for each output and plot the corresponding histogram over 1000 samples. The dash-line shows the Gaussian distribution of the same mean and standard deviation.}
\label{fig:hist}
\end{figure}

Since GSSN utilizes learned probability distributions to randomly sample its outputs, the resulting outputs possess a certain level of randomness. To provide an intuitive illustration of this randomness, we fix a particular problem and randomly sample the GSSN output $1000$ times, and Figure~\ref{fig:hist}  displays the histogram of feasible solution costs. Figure~\ref{fig:hist} reveals that the GSSN output approximately follows a Gaussian distribution regarding their cost function value. Given the relatively negligible sampling time (in our experimental setting, approximately 0.015 seconds per sample), we can readily augment the number of samples to optimize the search for feasible solutions that offer a lower computational cost.

\subsection{The Comparison of WarmUp Solutions }\label{sec:warmupmethods}

To evaluate the quality of feasible solutions generated by GSSN for the network flow problem in~\eqref{eq:milp_nonlin}, we conduct two sets of control experiments using the Gurobi solver and RSN algorithm, respectively. 

\begin{figure}[ht!]
\center
\begin{adjustwidth}{-0.2in}{-0.2in}
\includegraphics[width=0.99\linewidth, height = 0.5\linewidth]{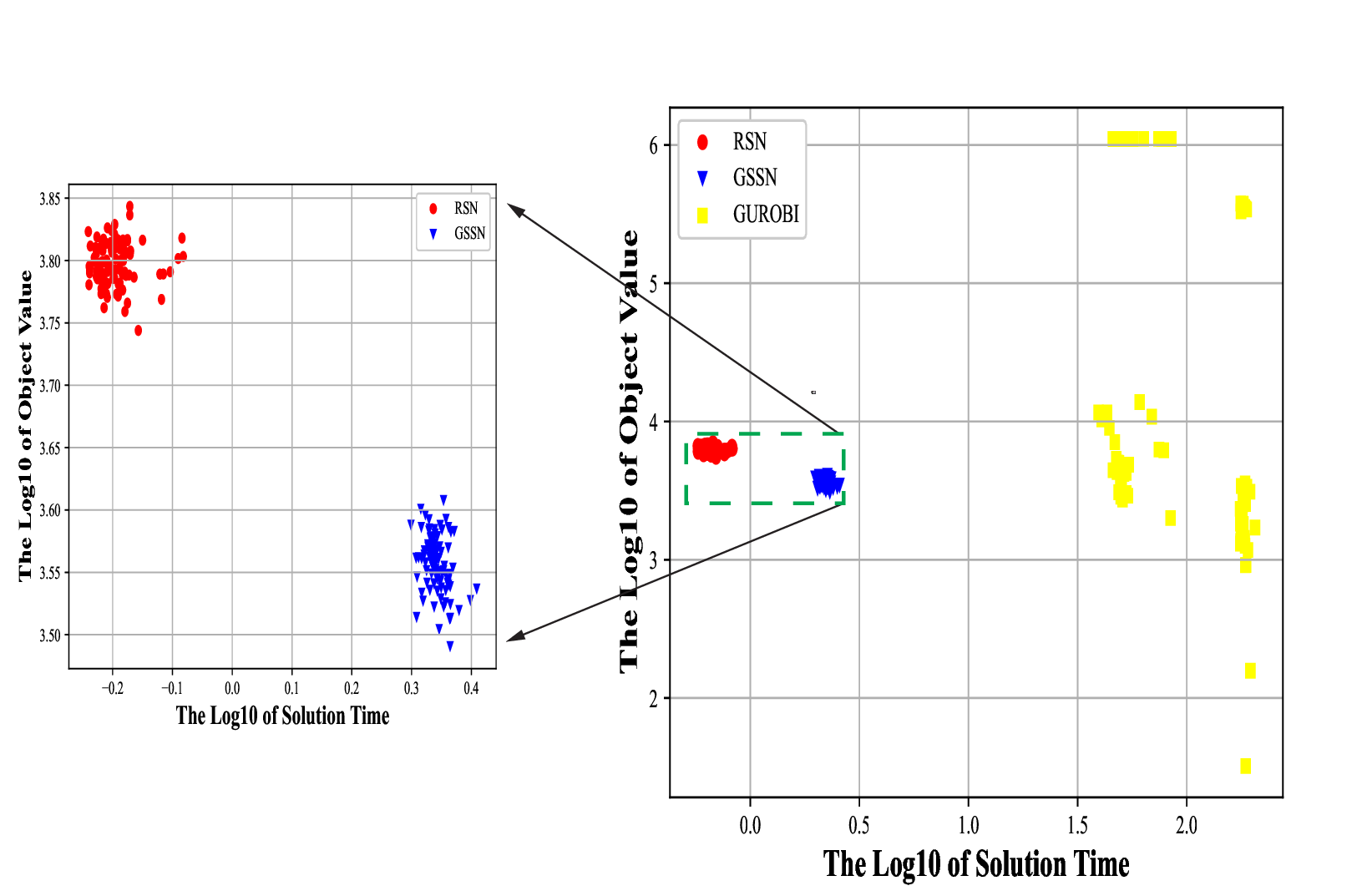}
\end{adjustwidth}
  \caption{The scatter plot of the solution time and corresponding objective function values for 100 test problems. In light of the substantial discrepancies in the scales of the three methods, a logarithmic scale with a base of $10$ is employed to construct the scatter plot.}
\label{fig:RGGScatter}
\end{figure}

We utilize three approaches to obtain initial feasible solutions for the problem. The first method involves configuring the Gurobi solver with a maximum search time of 300 seconds and employing the Gurobi code ``model.Params.SolutionLimit$=1$'' to prioritize the search for feasible solutions. The other two methods include utilizing the feasible solutions obtained through RSN and GSSN sampling for network flow problems and the initial feasible solution obtained through Gurobi computation. To provide an intuitive representation, a scatter plot (Figure~\ref{fig:RGGScatter}) is generated to illustrate the solution times and the corresponding objective function values (also known as cost) for the test problems.  The mean and standard deviation (std) of the objective function values and solution time for the three methods can be found in Table~\ref{tab:comparison}. {\color{black} It is worth mentioning that due to the assumption of capacity parameters in \eqref{eq:assumpDemand}, the relaxation solutions of all the generated test problems result in zero cost function values, leading to a trivial optimal gap\cite{optimalgap} curve during the solving process. Instead, we will utilize the cost function value to assess the performance of different methods.
}

\begin{table}[htbp]
\def\arraystretch{1.25}
    \centering
    \caption{Mean and standard deviation of cost and solution time}
    \label{tab:comparison}
    \begin{tabular}{|c|c|c|c|c|c|c|}
        \hline
        \multicolumn{2}{|c|}{\multirow{2}*{Statistics}} & \multicolumn{2}{c|}{Cost} &\multicolumn{2}{c|}{Time}\\
        \cline{3-6}
        \multicolumn{2}{|c|}{~}&mean&std &mean &std \\
        \hline
        \multicolumn{2}{|c|}{RSN}&6272.1357 & 248.8282 &\textbf{\textcolor{black}{0.6403}} &\textbf{\textcolor{black}{0.0514}}\\
        \hline
         \multicolumn{2}{|c|}{GSSN}&\textbf{\textcolor{black}{3607.3104}} &\textbf{\textcolor{black}{181.9166}} &2.198760&0.0922\\
         \hline
          \multicolumn{2}{|c|}{Gurobi}&476182.95 &529995.43 &95.0689 & 61.4607\\
        \hline
    \end{tabular}
\end{table}

Based on the analysis of Figure~\ref{fig:RGGScatter} and Table~\ref{tab:comparison},  noticeable differences are observed among the three methods in terms of the time required to obtain feasible solutions, where RSN exhibits the shortest time, followed by GSSN, and then Gurobi with a significant margin. Regarding the computed feasible solution values, GSSN produces the smallest objective function values, followed by RSN, then Gurobi.

Interestingly, approximately half of the feasible solutions obtained by Gurobi have significantly larger magnitudes, reaching up to 1e6, compared to the other half, which has magnitudes similar to those obtained by GSSN and RSN. When comparing GSSN with RSN, the mean cost obtained by RSN is around twice that of GSSN. Besides, GSSN has a smaller standard deviation, indicating a more concentrated and stable distribution of feasible solutions.

\subsection{Use Gurobi to Test the Solution Quality of Three Warmups} \label{sec:test_warm}

\begin{figure}[htbp]
\begin{adjustwidth}{-0.0in}{0in}
    \centering
{\mbox{\includegraphics[width=.7\linewidth]{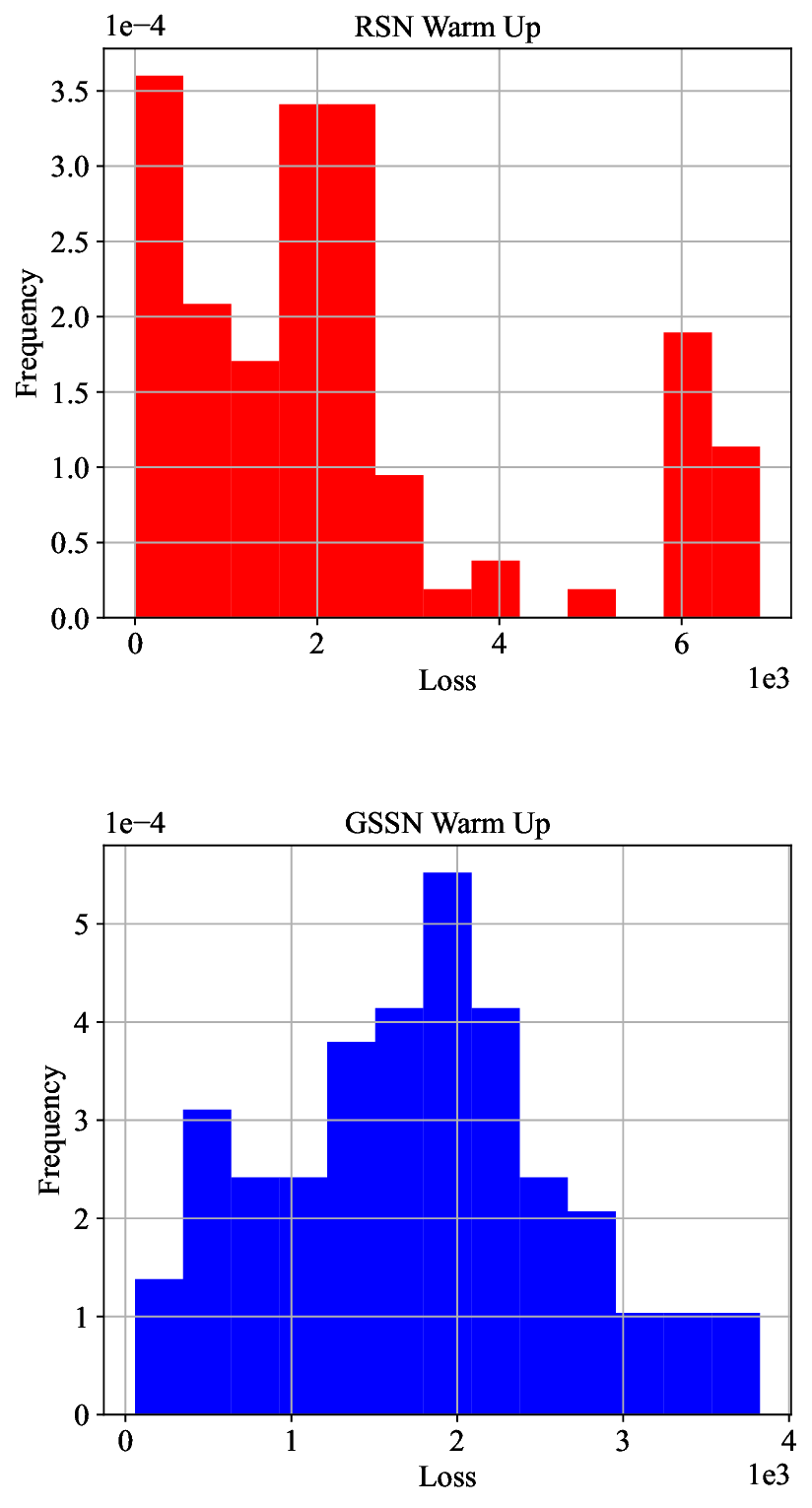}}} {\mbox{\includegraphics[width=.8\linewidth]{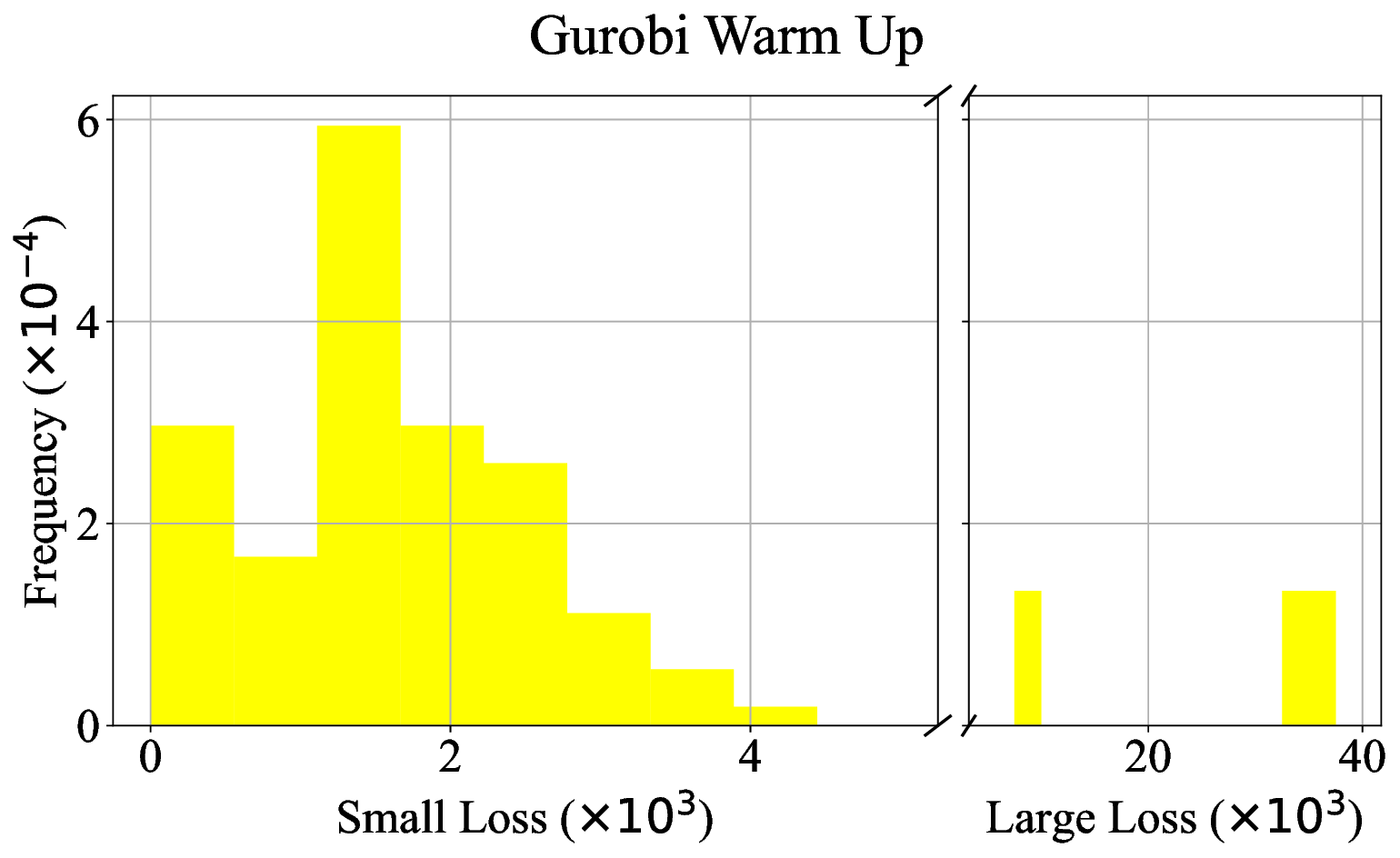}}}
\end{adjustwidth}
  \caption{The histogram of the cost at the end of the solver's 300-second runtime. The subplots from top to bottom correspond to the RSN warmup, GSSN warmup, and Gurobi warmup, respectively. {\color{black} (We used a non-uniform axis spacing in the historgram for Gurobi warmup to reveal the details at the small-loss region)} In accordance with the three feasible solution generation methods discussed in Section~\ref{sec:warmupmethods}, we employed the obtained feasible solutions as initial conditions for Gurobi.}
\label{fig:WarmUp5}
\end{figure}

{\color{black} We utilize the feasible solutions generated by the three methods introduced in Section~\ref{sec:warmupmethods} as warm start inputs for Gurobi for the 100 test problems in Section~\ref{sec:setting}. A maximum solving time of 300 seconds is set for each problem. (See Appendix~\ref{app:cost} for the details) The experiment aims to evaluate the impact of these three potential solutions on Gurobi's short-term solving capacity to simulate real-world network scheduling scenarios.} The cost of the problems solved by the three methods after 300 seconds is recorded. The histograms of the cost distributions obtained by the three methods as warm-up are shown in Figure~\ref{fig:WarmUp5}, and the corresponding statistics of the mean and standard deviation of the costs can be found in Table~\ref{tab:comparisonWarmUp}.

\begin{table}[htbp]
\def\arraystretch{1.25}
    \centering
    \caption{Mean and standard deviation of cost based on three warmups}
    \label{tab:comparisonWarmUp}
    \begin{tabular}{|c|c|c|c|}
        \hline
        \multicolumn{2}{|c|}{\multirow{2}*{Statistics}} & \multicolumn{2}{ c|}{Cost} \\
        \cline{3-4}
        \multicolumn{2}{|c|}{~}&mean&std \\
        \hline
        \multicolumn{2}{|c|}{RSN}&2300.1284 &1979.9320  \\
        \hline
         \multicolumn{2}{|c|}{GSSN}&\textbf{\textcolor{black}{1751.6571}} &\textbf{\textcolor{black}{852.9082}} \\
         \hline
          \multicolumn{2}{|c|}{Gurobi}&2344.7750 &4882.5962 \\
        \hline
    \end{tabular}
\end{table}

From Figure~\ref{fig:WarmUp5} and Table~\ref{tab:comparisonWarmUp}, we observe that the GSSN method has the smallest mean and standard deviation among the three feasible solution generation methods. This indicates that the warmup feasible solutions generated by GSSN are of superior quality and more likely to escape the attraction domain of local minima with high-cost values. {\color{black} Thus, we can implement the GSSN as a pre-conditioner for classical solvers like Gurobi.}

\subsection{Generalization Test of GSSN}

Given our objective of evaluating the generalization capabilities of the GSSN and RSN sampling algorithms in dynamic scenarios, where rapid generation of feasible solutions and traffic allocation plans is essential, our focus is primarily on changes in problem cost values. Our experiments reveal that Gurobi’s performance in finding feasible solutions deteriorate for user counts exceeding 15, with some problems requiring more than 300 seconds to find a feasible solution. Consequently, we do not employ Gurobi to obtain more accurate solutions or assess the quality of longer-duration solutions. As such, Gurobi is excluded from subsequent experimental comparisons.

The training set for GSSN is constructed by simulating a scenario involving 10 users and 48 time slots for problem~\eqref{eq:milp_nonlin}. Subsequently, we aim to evaluate the generalizability of the trained GSSN model in scenarios characterized by more users and more time slots. The specific experimental configurations for assessing the model's generalization performance are provided below:

\begin{itemize}
    \item \textbf{the generalization of time slots}:  We fix the number of users ($N_e = 10$) and static parameters for problem~\eqref{eq:milp_nonlin}. We generate 100 independent and identically distributed problems for each time slot size by sampling the inbound/outbound traffic demand.
    \item \textbf{the generalization of user numbers}: We fix the number of time steps at 48 for network flow problem \eqref{eq:milp_nonlin} and randomly generate 100 problems for each user number. In contrast to the parameter generation method used in training data generation, the 100 problems are generated by \textbf{independently and identically sampling both the static and dynamic parameters}. 
\end{itemize}

\begin{figure}[ht!]
\center
\begin{adjustwidth}{-0.2in}{-0.2in}
\includegraphics[width=0.49\linewidth]{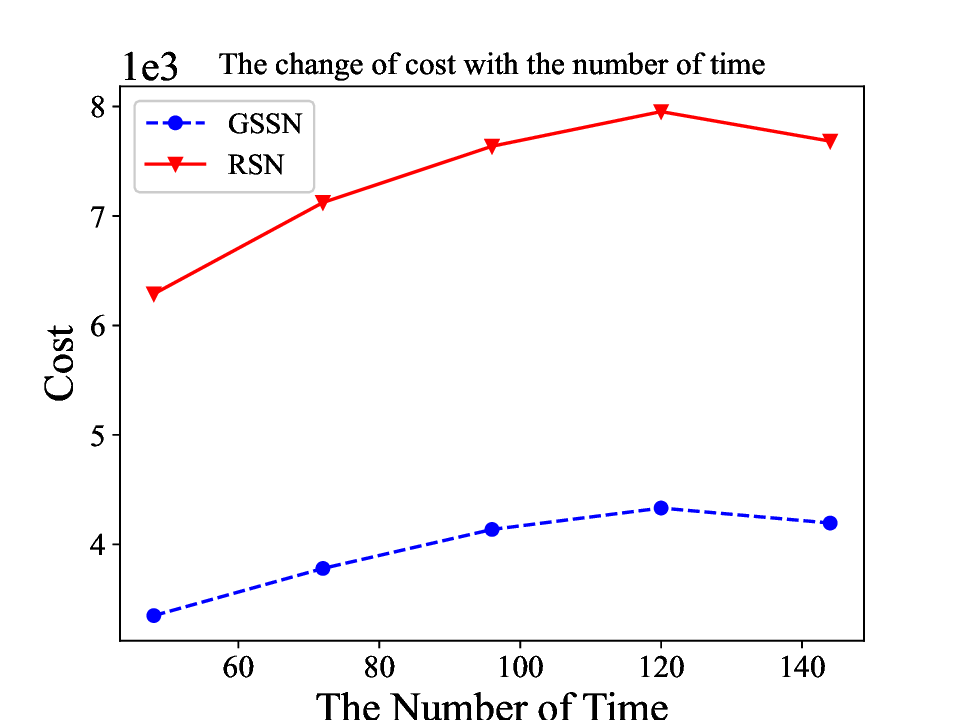}
\includegraphics[width=0.49\linewidth]{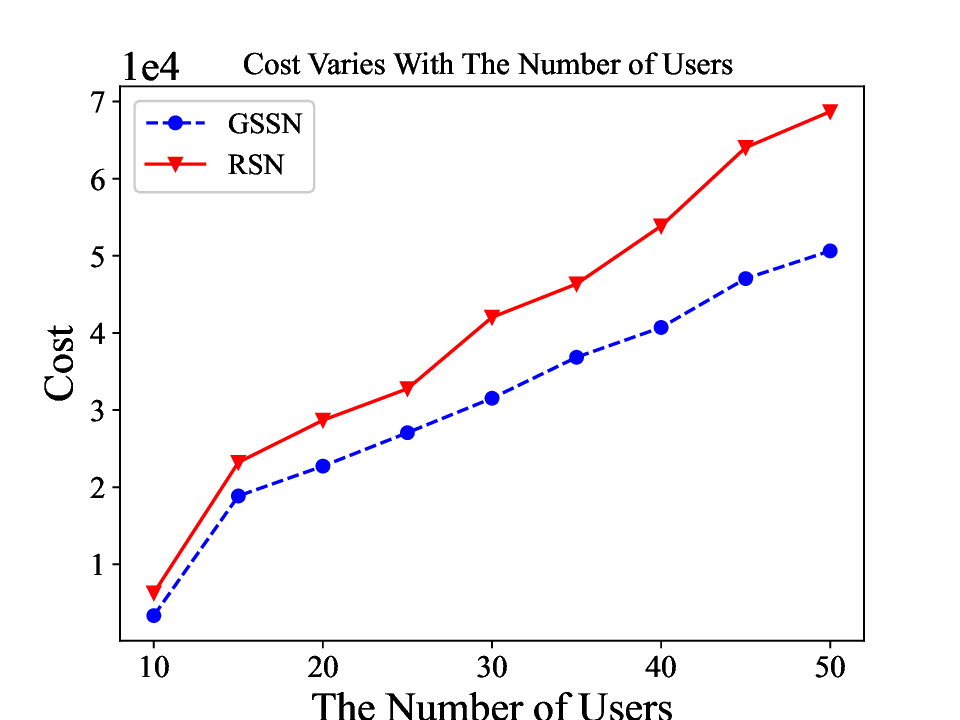}
\end{adjustwidth}
\caption{Generalization tests of the trained neural network on the number of
time slots(left panel) and the number of users(right panel).}
\label{fig:timegeneralize}
\end{figure}

Next, we compute the average cost of the generated feasible solutions for these 100 problems in the GSSN and RSN models. We graph the average costs obtained in relation to the number of time slots or users, as illustrated in Figure~\ref{fig:timegeneralize}. We can see that the average cost of GSSN and RSN sampling increases approximately linearly as the number of time slots and users increases. However, the GSSN consistently outperforms the RSN model in generating feasible solutions at lower costs. 

% During training, we fix the static parameters and generate data by varying users' inbound and outbound traffic demand. In real-world scenarios, network flow problems are subject to dynamic changes in network static parameters and traffic demand. 

\begin{figure}[ht!]
\begin{adjustwidth}{-0.2in}{-0.2in}
\center
  \includegraphics[width=0.49\linewidth]{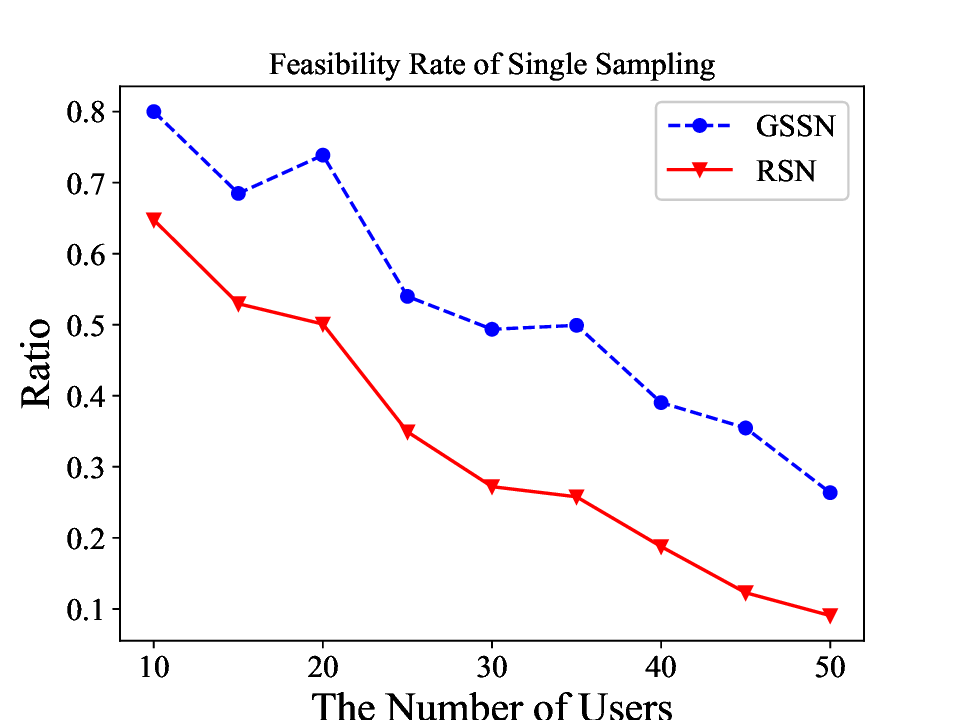}
  \includegraphics[width=0.49\linewidth]{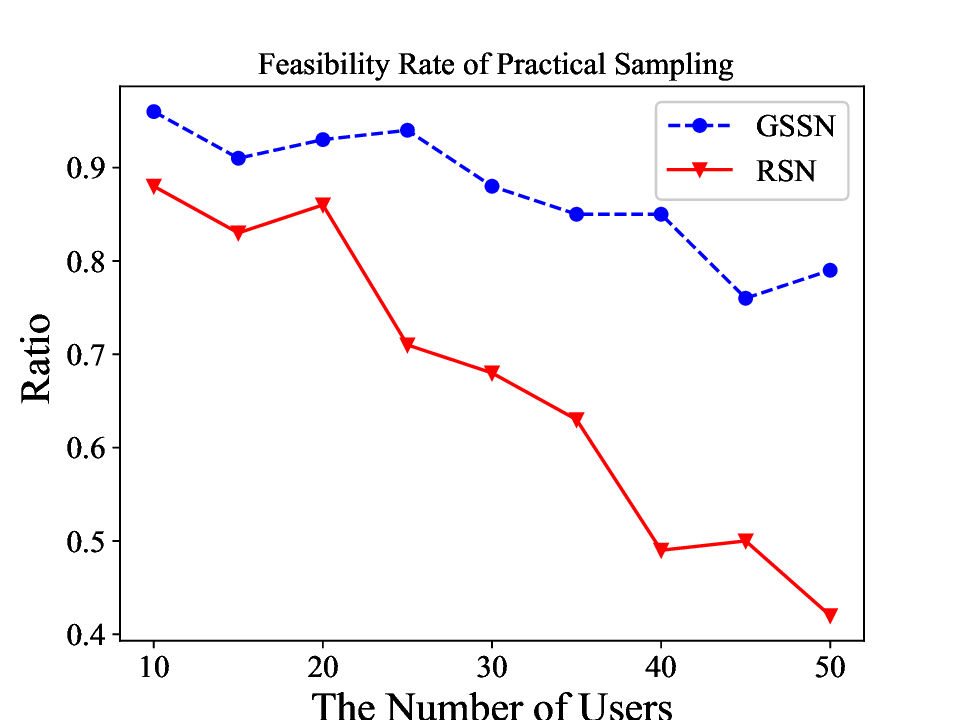}
\end{adjustwidth}
  \caption{The SSFR(left panel) and PFR(right panel) for GSSN (blue-dash) and RSN(red-solid) subject to the generalization test on user numbers}
\label{fig:FeasibleSampleOnegeneralize}

\end{figure}

% To assess the generalization performance of GSSN in dynamic scenarios, we fix the number of time steps at 48 for network flow problem \eqref{eq:milp_nonlin} and randomly generate 100 problems for each user numbers. In contrast to the parameter generation method used in training data generation, the 100 problems are generated by \textbf{independently and identically sampling both the static and dynamic parameters}. 
In the generalization experiments involving time slots, both models demonstrate consistent success in sampling feasible solutions for each test problem. However, in the generalization experiments pertaining to varying user numbers, it is noteworthy that the success rate of generating feasible solutions by both GSSN and RSN models did not reach 100\%. To assess the level of difficulty in generating feasible solutions through GSSN and RSN sampling, we establish two metrics. 
\begin{itemize}
    \item \textbf{the single sampling feasibility rate of the algorithm(SSFR):} For a given set of $N$ problems, each problem is sampled $M$ times, and the number of successful samples among $N*M$ samples is denoted as $L$. The single sampling feasibility rate is then calculated as $SSFR = L/(MN)$. 
    \item \textbf{the feasibility rate of the practical algorithm(PFR):} For a given set of $N$ problems, each problem is sampled $M$ times. If at least one feasible solution is generated among $M$ samples, then the number of problems that generate feasible solutions among $N$ problems is denoted as $S$. The feasibility rate of the practical algorithm is then calculated as $PFR = S/N$.
\end{itemize}
Figure~\ref{fig:FeasibleSampleOnegeneralize} show the value of SSFR and PFR for different numbers of users.

Further examination of Figure~\ref{fig:FeasibleSampleOnegeneralize} shows that as the number of users and problem difficulty increase, the feasibility rates of both GSSN and RSN decline. However, GSSN exhibits a higher probability of generating a feasible solution with a single sample than RSN. When employing an algorithm that samples 100 times and evaluating the PFR indicator, we observe that while the feasibility rates of both GSSN and RSN decrease with increasing numbers of users in Figure~\ref{fig:FeasibleSampleOnegeneralize}, the ratio of decline for GSSN is significantly slower than that of RSN. For user counts ranging from 10-50, the feasibility rate of GSSN remains above $70\%$, whereas for user counts exceeding 25, the feasibility rate of RSN falls below $70\%$. \textcolor{black}{We want to clarify that in some application scenarios, the number of users may be as low as 10-50. For example, in Appendix A.4 of \cite{Singh21}, the actual Cloud WAN considered has 50 users. Also, many Cloud WAN setups within China, represented by Huawei Cloud and Alibaba Cloud, rely on BGP links to establish connectivity with their cloud services across 34 provinces and districts, which leads to edge-cloud networking of 34 users. Therefore, our model has a practical application to a certain extent.}

\section{Conclusion and Discussion}
\label{sec:conclusion}

We formulated the quantile optimization problem for edge-cloud traffic scheduling into constraint integer optimization problems with parameters describing traffic demands and link capacities. For the edge-cloud networking problem, we solved the corresponding unconstraint continuous optimization problem via unsupervised learning using GSSN, where the inputs and outputs are the problem parameters and the allocation scheme, respectively. In the numerical experiments, we tested the quality of the outputs regarding the objective function value, warmups for classical solvers like Gurobi, and the generalization property by varying the number of time slots and users. From the numerical results, the GSSN method outperforms classical solvers and the RSN approach, and the output acts as short-term warmups for Gurobi. The GSSN shows a reasonable generalization property essential in real-world applications.

Regarding future works, so far, we have only tested the capability of GSSN in solving the scheduling problem in the SD-WAN environment with a relatively small number of users and time slots in a billing cycle. To better evaluate the performance of our approach, we plan to extend the GSSN to more general edge-cloud networking situations that include a larger number of users and a more complex billing scheme. The proposed problems are within the scope of a static regime, where we assume we know the traffic demand in advance. However, the allocation of traffic depends on real-time traffic demands, making the dynamic regime more significant in practice. 

Also, despite the good numerical outcomes, using the soft-loss function as the objective function in training does not theoretically guarantee the output is always feasible. We already observed the decay of the feasible rate in the generalization test. Improving the feasible rate for neural network-based methods remains a critical issue in solving complex integer programming problems, and the GSSN is no exception. In the future, we need to develop more systematic tools to ensure the feasibility of the GSSN output and seek opportunities to implement GSSN on more challenging edge-cloud networking problems.

\appendices

\section{The Problem Data Generating Method} \label{app:DataGeneration}

The current appendix delineates the methodology for generating training and test sets used in Section~\ref{sec:num}. Following the interpretation in Remark~\ref{rk:3} and simplicity reasons, we propose the following assumptions regarding the problem set and parameters therein:
\begin{itemize}
    \item the problem set is categorized by its network topology (Figure~\ref{fig:network}), where the problems share the same static parameters (Table~\ref{tab:2}) values;
    \item the physical maximum link capacities ($c_e^M, c_{\ell}^M$) should be large enough such that the related constraints in \eqref{eq:milp_nonlin} hold constantly;
    \item for each problem, the inbound and outbound traffic demands at different time slots $\{\bar{d}_{k,n}^{t},  \underline{d}_{k,n}^{t}\}_{t=1}^{T}$ are i.i.d. samples for each $k$ and $n$.
\end{itemize}
Since we still ask the billable bandwidth to be no greater than the maximum link capacities ($c_{e}^{m}$), truncating the constraints regarding the physical maximum link capacities ($c_{e}^{M}$) does not intrinsically change the problem complexity. Meanwhile, such truncations help us tune the parameter value so that the difficulty of finding a feasible solution by RSN is reasonable. Guided by the assumptions, we initiate the static parameter values and introduce the sampling method, subject to the existing problem data, for dynamic parameters afterward.

\begin{table}[ht!]
\def\arraystretch{1.25}
    \centering
    \caption{Generating some of the static parameter values}
    \label{tab:basic}
    \begin{tabular}{|p{1.4cm}<{\centering}|p{2.8cm}|p{3.45cm}<{\centering}|}
        \hline 
        Symbol & Meaning & Value\\  
         \hline
        {$N_{e}$} & {the number of edges/users} & 10 \\
        \hline
        {$T$} & the number of time slot & 48 \\
        \hline
        {$K$} & the number of traffic demand types & 8\\
        \hline
        {$N_{I}$} & the number of ISPs & 4\\
        \hline
        {$c_e^M$} & the physical maximum link capacity & 10000\\
        \hline
        {$c_e^m$} & the maximum link capacity & $ \sim \mathrm{Uniform}(300, 1000)$\\
        \hline
        {$c_e^b$} & the basic link capacity & $\sim \mathrm{Uniform}(0.05*c_e^m, 0.5*c_e^m)$\\
        \hline
        {$r_e$} & the peering link rate & $ \sim\mathrm{Uniform}(5, 10)$\\
        \hline
        {$E_{k,n}$} & the set of admissible peering link & $ \sim\mathrm{Binomial}(EL,0.5)$, if empty set $E_{k,n} = E_{n}$ \\
        \hline
        $\tau_{start},\tau_{end}$ & the hyperparameters of the Gumbel-Softmax& 2,\; 0.31\\
        \hline
        $N_{epochs}$& the maximum training epoch & 100\\
        \hline
    \end{tabular}
\end{table}

\subsection*{Static Parameter Values}

A portion of the static parameter values is determined directly as illustrated in Table~\ref{tab:basic}. Here, $\mathrm{Uniform}(a,b)$ and $\mathrm{Binomial}(N_{I},0.5)$ stand for the uniform distribution on interval $[a,b]$ and binomial distribution of $N_{I}$ Bernoulli trials, respectively.

The remaining static parameters are the link capacities and link rates of the ISP links, which should be consistent with the parameter value of the edge links connected to the ISP. Since the ISP merges all the edge link traffic connecting them (Figure~\ref{fig:network}), the ISP link capacities should correspond to the total link capacities of the edge links connected. For example, the specific generation method of $c_{\ell_i}^{b}$ adheres to
\begin{equation}\label{eq:hub}
c^{b}_{\ell_i} \sim \mathrm{Uniform}(0.8*C^{b}_{\ell_i},0.9*C^{b}_{\ell_i}),\; C^{b}_{\ell_i} = \sum_{n=1}^{N} c^{b}_{e_{n,i}},
\end{equation}
where $i = 1,2,3,4$. The rest ISP link capacities $c^{m}_{\ell_{i}}$ and $c^{M}_{\ell_i}$ are sampled in the same manner as Eq.~\eqref{eq:hub}. In \eqref{eq:hub}, we introduced a pair of contraction coefficients that can be adjusted to produce problem sets of desirable difficulties.

\subsection*{Dynamic Parameter Values}

Subsequently, subject to fixed static parameter values, we generate samples of the dynamic parameters, i.e., the traffic demands, which lead to the training and test problem sets used in Section~\ref{sec:num}. Introducing the inbound and outbound traffic demands random tensors,
\begin{equation}\label{eq:dem_mat}
    \bar{D} = (\bar{d}^{t}_{k,n})\in \mathbb{R}^{K\times N_{e} \times T}, \; \underline{D} = (\bar{d}^{t}_{k,n})\in \mathbb{R}^{K\times N_{e} \times T},
\end{equation}
respectively. Motivated by the assumption, we establish Algorithm~\ref{alg:demand} to efficiently generate i.i.d. samples of the element $\bar{d}^{t}_{k,n}$ and $\underline{d}^{t}_{k,n}$.

\begin{algorithm}[ht!]
\caption{Sample inbound and outbound traffic demands} \label{alg:demand}
\begin{algorithmic}
\STATE \textbf{Input:} the static parameter values (Table~\ref{tab:basic})
\STATE Initiate by 
\begin{equation*}
\bar{d}^{t}_{k,n},\underline{d}^{t}_{k,n} \sim \mathrm{Uniform}(20,30)
\end{equation*}
\STATE  \textbf{for} $n = 1,2,\dots,N_{e}$:
\STATE \hspace{.5cm} Compute
\begin{equation*}
    cb = \sum_{e\in E_{n}} c^b_{e},\quad cm = \sum_{e \in E_{n}} c^m_{e}
\end{equation*}
\STATE \hspace{.5cm} {\textbf{for}} $t = 1,2,\dots,T$:
\STATE \hspace{1.0cm} Compute
\begin{equation*}
\bar{as} = \sum_{k=1}^K \bar{d}^{t}_{k,n},\quad
\underline{as}  = \sum_{k=1}^K \underline{d}^{t}_{k,n} 
\end{equation*}
\STATE \hspace{1.0cm} Sample a scalar $p\sim \mathrm{Uniform}(0,1)$ 
\STATE  \hspace{1.0cm} \textbf{if}   $p<0.5$: update $(\bar{D}(:,n,t),\underline{D}(:,n,t))$ by
\begin{equation}\label{eq:low}
    \begin{split}
    &\bar{d}^{t}_{k,n} \sim \mathrm{Uniform}\left(0.6*\frac{\bar{d}^{t}_{k,n}*cb}{\bar{as}}, \; 0.8*\frac{\bar{d}^{t}_{k,n}*cb}{\bar{as}}\right)\\
    &\underline{d}^{t}_{k,n} \sim \mathrm{Uniform}\left(0.6*\frac{\underline{d}^{t}_{k,n}*cb}{\underline{as}},\; 0.8*\frac{\underline{d}^{t}_{k,n}*cb}{\underline{as}} \right)
    \end{split}
\end{equation}
\STATE  \hspace{1.0cm} \textbf{else}: update $(\bar{D}(:,n,t),\underline{D}(:,n,t))$ by
\begin{equation}\label{eq:high}
    \begin{split}
    &\bar{d}^{t}_{k,n} {\sim} \mathrm{Uniform}\left(0.6*\frac{\bar{d}^{t}_{k,n}*cm}{\bar{as}}, 0.8*\frac{\bar{d}^{t}_{k,n}*cm}{\bar{as}}\right) \\
    &\underline{d}^{t}_{k,n} {\sim} \mathrm{Uniform}\left(0.6*\frac{\underline{d}^{t}_{k,n}*cm}{\underline{as}}, 0.8*\frac{\underline{d}^{t}_{k,n}*cm}{\underline{as}} \right)
    \end{split}
\end{equation}
\STATE \hspace{1.0cm} {\textbf{end}}
\STATE \hspace{.5cm} {\textbf{end}}
\STATE \hspace{.5cm} Compute
\begin{equation*}
    cbb = \sum_{e\in E_{n}} c^b_{e}
\end{equation*}
\STATE \hspace{.5cm} \textbf{for} $k = 1,2,\dots,K$:
\STATE \hspace{1.0cm} Update $(\bar{D}(k,n,:), \underline{D}(k,n,:))$ by
\STATE\hspace{1.0cm} {\textbf{if}} $\bar{d}^{t}_{k,n} > 0.25*cbb$:
\STATE\hspace{1.0cm} 
\begin{equation}\label{eq:up}
\bar{d}^{t}_{k,n} \sim \mathrm{Uniform}(0.05*cbb,0.125*cbb)
\end{equation}
\STATE\hspace{1.0cm} {\textbf{end}}
\STATE\hspace{1.0cm} {\textbf{if}} $\underline{d}^{t}_{k,n} > 0.25*cbb$:
\STATE\hspace{1.5cm} 
\begin{equation}\label{eq:down}
    \underline{d}^{t}_{k,n} \sim \mathrm{Uniform}(0.05*cbb,0.125*cbb)
\end{equation}
\STATE\hspace{1.0cm} {\textbf{end}}
\STATE \hspace{0.5cm} {\textbf{end}}
\STATE \textbf{end}
\STATE \textbf{return} The traffic demand random tensors $(\bar{D},\underline{D})$.
\end{algorithmic}
\end{algorithm}

\begin{figure}[ht!]
\center

\includegraphics[width=0.75\linewidth]{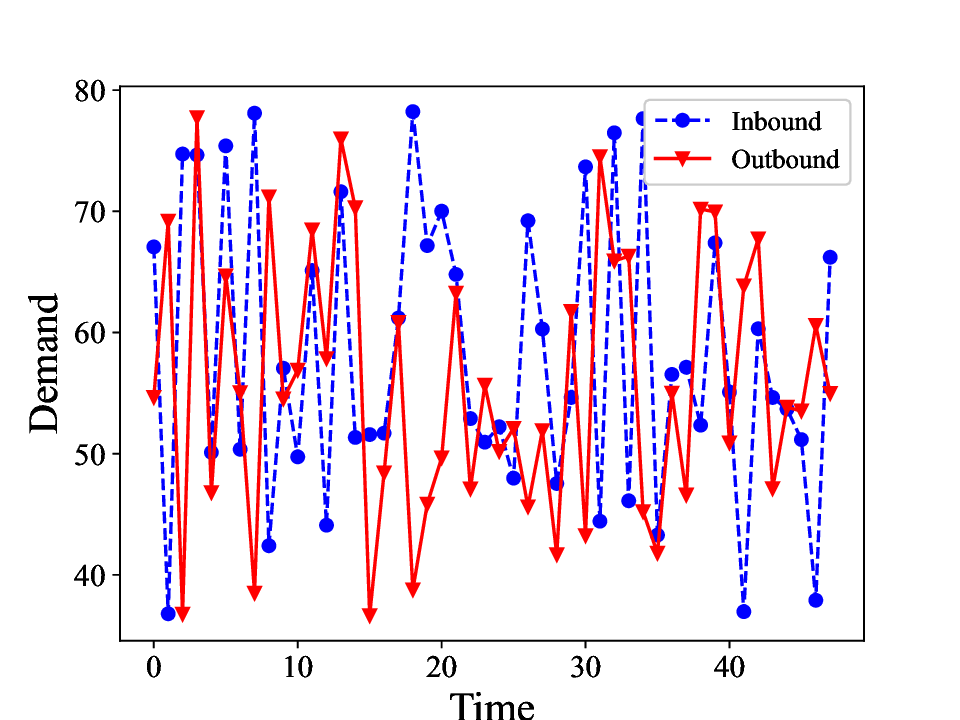}
  \caption{An example of the inbound/outbound traffic demand in the randomly generated test problems. The figure shows an example of the traffic demand, randomly generated to simulate the rapid change of the demand.}
\label{fig:demand}
\end{figure}

To remove possible extreme traffic demand, we employ Eqs. \eqref{eq:up} and \eqref{eq:down} to ensure that the aggregate inbound and outbound traffic demands of user $n$ at time $t$ are controlled by the sum of the maximum link capacity of all connected edges. Take the inbound traffic demands as an example, and we have
\begin{equation*}
\sum_{k=1}^{K} \bar{d}^{t}_{k,n} \leq \sum_{k=1}^{8} 0.25*cbb = 2\sum_{e\in E_{n}} c^b_{e} \leq \sum_{e\in E_{n}} c^m_{e}.
\end{equation*}
In Algorithm~\ref{alg:demand}, the computation in the $t$-loop and $k$-loop can be processed via vector operations, which benefits the efficiency. Figure~\ref{fig:demand} depicts an example of inbound/outbound traffic demand sampling in a scenario with a single user.

\section{The Linearized Network Scheduling Problem} \label{app:linear}

In this appendix, we introduce the linearization of the network scheduling problem in \eqref{eq:milp_nonlin}. To resolve the nonlinearity introduced by the quantile function $g_{95}$, we introduce 0-1 intermediate variables $(\bar{u}_{e,n}^{t}, \underline{u}_{e,n}^{t})$, $e\in E_n$, $n=1,2,\dots, N_{e}$ and $(\bar{u}_{\ell_i}^{t}, \underline{u}_{\ell_{i}}^{t})$, $i=1,2,3,4$ to label the top $5\%$ inbound/outbound traffic time slots for edge links and ISP links, respectively. For example, $\bar{u}_{e,n}^{t}=1$ means the averaged inbound traffic of the current time slot belongs to the top $5\%$ inbound traffic among the billing period. We propose the following linearized constraint for the problem \eqref{eq:milp_nonlin}.

 %     \item  the edge link allocation should meet the traffic demand and follow the split ratio
 %     \begin{equation*}
 %         \begin{split}
 %     &\sum_{e\in E_{k,n}} y_{e,k,n}^{t} \geq 1, \quad \forall k,n,t; \\
 %     &\sum_{e\in E_{k,n}}\bar{x}_{e,k,n}^{t} = \bar{d}_{k,n}^{t}, \quad \sum_{e\in E_{k,n}}\underline{x}_{e,k,n}^{t} = \underline{d}_{k,n}^{t}, \\
 % & c_{e_2}^{b} \bar{x}_{e_1,k,n}^{t} \geq c_{e_1}^{b} \bar{x}_{e_2,k,n}^{t} + M\cdot( y_{e_1,k,n}^{t} + y_{e_2,k,n}^{t} - 2), \\
 % & c_{e_2}^{b} \underline{x}_{e_1,k,n}^{t} \geq c_{e_1}^{b} \underline{x}_{e_2,k,n}^{t} + M\cdot( y_{e_1,k,n}^{t} + y_{e_2,k,n}^{t} - 2),  \\
 % &\forall e_{1}\not=e_{2} \in E_{k,n}, \quad \forall k,n,t; \\
 %         \end{split}
 %    \end{equation*}

\begin{enumerate}
\item identify the billable bandwidth of the edge links
 \begin{equation*}
      \begin{split}
 &\sum_{t=1}^{T} \bar{u}_{e,n}^{t} \leq 0.05T, \;  \sum_{t=1}^{T} \underline{u}_{e,n}^{t} \leq 0.05T, \quad \forall e\in E_n, \; \forall n, \\ 
 & \bar{f}^{t}_{e,n} = \sum_{k=1}^{8}  \bar{x}^{t}_{e, k,n}, \quad \underline{f}_{e,n}^{t} =  \sum_{k=1}^{8}  \underline{x}^{t}_{e,k,n}, \\
 & z_{e} \geq  \bar{f}_{e,n}^{t} - c_{e}^{M}\bar{u}_{e,n}^{t}, \\
 & z_{e} \geq \underline{f}_{e,n}^{t} - c_{e}^{M}\underline{u}_{e,n}^{t}, \quad \forall t, \;\forall e\in E_n,\; \forall n;
      \end{split}
  \end{equation*}

 \item identify the billable bandwidth of the ISP links
  \begin{equation*}
      \begin{split}
 &\sum_{t=1}^{T} \bar{u}_{\ell_i}^{t} \leq 0.05T, \;  \sum_{t=1}^{T} \underline{u}_{\ell_i}^{t} \leq 0.05T, \quad i=1,2,3,4; \\
 &\bar{X}_{\ell_{i}}^{t} = \sum_{n=1}^{N_{e}} \bar{f}^{t}_{e_{n,i},n}, \quad \underline{X}_{\ell_{i}}^{t} = \sum_{n=1}^{N_{e}}  \underline{f}_{e_{n,i},n}^{t}, \\
 & z_{\ell_i} \geq \bar{X}_{\ell_i}^{t} - c_{\ell_i}^{M}\bar{u}_{\ell_i}^{t},\\
 & z_{e} \geq \underline{X}_{\ell_i}^{t} - c_{\ell_i}^{M}\underline{u}_{\ell_i}^{t}, \quad \forall t, \; i=1,2,3,4; \\
      \end{split}
  \end{equation*}
  \item the link capacities constraint
  \begin{equation*}
      \begin{split}
 & \bar{f}_{e,n}^{t} \leq c_{e}^{m}\cdot (1 - \bar{u}_{e,n}^{t}) + c_{e}^{M}\cdot \bar{u}_{e,n}^{t}, \\
 & \underline{f}_{e,n}^{t} \leq c_{e}^{m}\cdot (1 - \underline{u}_{e,n}^{t}) + c_{e}^{M}\cdot \underline{u}_{e,n}^{t}, \\
 & \forall t, \; \forall e\in E_{n}, \; \forall n \\
 & \bar{X}_{\ell_i}^{t} \leq c_{\ell_i}^{m}\cdot (1 - \bar{u}_{\ell_i}^{t}) + c_{\ell_i}^{M}\cdot \bar{u}_{\ell_i}^{t}, \\
 & \underline{X}_{\ell_i}^{t} \leq c_{\ell_i}^{m}\cdot (1 - \underline{u}_{\ell_i}^{t}) + c_{\ell_i}^{M}\cdot \underline{u}_{\ell_i}^{t}, \quad \forall t,\; i =1,2,3,4;\\
  & z_{e} \leq c_{e}^{m}, \quad \forall e\in E, \\
 & z_{\ell_i} \leq c_{\ell_i}^{m}, \quad  i =1,2,3,4. \\
      \end{split}
  \end{equation*}
\end{enumerate}

{\color{black}
\section{The decay of cost functions (timeout: 6 hours)} \label{app:cost}

This appendix reports the long-time behavior of the cost function during Gurobi solving process. We randomly selected 4 problems from the test problems used in Section~\ref{sec:test_warm} and plotted the trajectories of the resulting cost functions in Figure~\ref{fig:cost_fun}. The cost functions' decay behavior indicates that after the decline in the first few minutes,  it takes a significant amount of time to reach the next better solution, regardless of the choice of the initial conditions. Moreover, considering the online feature of the network scheduling task in practice, we employed a 300-second maximum solving time limit in the numerical experiments in Section~\ref{sec:test_warm}.

\begin{figure}[ht!]
        \centering
        \begin{adjustwidth}{-0.1in}{-0.1in}    \includegraphics[width=0.99\linewidth]{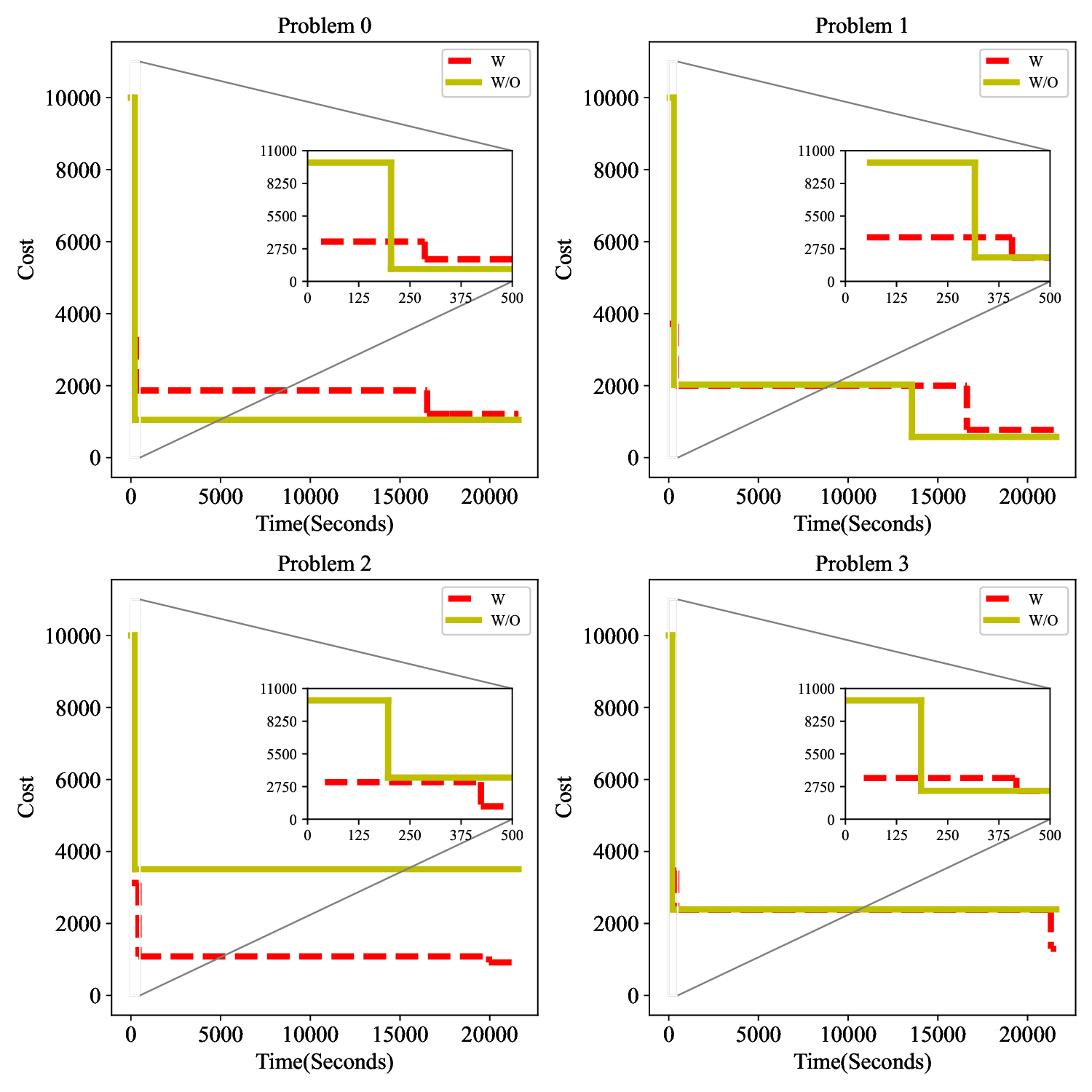}
        \end{adjustwidth}
        \caption{The trajectories of the cost functions of 4 test problems. Gurobi repeatedly solved each problem without a warm start (yellow-solid lines)  and with a warm start generated by GSSN (red-dash lines) for 6 hours to find solutions. We tracked the cost over time. At the beginning of the solving process, we manually marked the cost value as 10000 before the decision variable reached the feasible domain.}
        \label{fig:cost_fun}
    \end{figure}

}

\bibliographystyle{IEEEtran}
\bibliography{ref}

\end{document}